\documentclass[12pt,twoside, a4paper]{article}
\pdfoutput=1
\def\pd{\partial}
\def\mc{\mathcal}
\def\ul{\underline}

\usepackage[dvips]{graphicx}
\usepackage{amssymb}
\usepackage{amssymb,amsmath}
\usepackage{graphicx}
\usepackage{dsfont}
\usepackage{caption}
\usepackage{subcaption}
\usepackage{mathtools}
\usepackage{verbatim}
\usepackage{graphicx}
\usepackage{multirow}
\usepackage[outline]{contour}
\usepackage{xcolor,colortbl}
\usepackage{pdflscape}
\input{epsf.sty}  
\contourlength{.1pt}
\graphicspath{ {./Figures/} }
\numberwithin{equation}{section}

\begin{document}
\begin{center}
\Large{\textbf{Supersymmetric domain walls in maximal 6D gauged supergravity III}}
\end{center}
\vspace{1 cm}
\begin{center}
\large{\textbf{Patharadanai Nuchino}$^a$ and \textbf{Parinya Karndumri}$^b$}
\end{center}
\begin{center}
$^a$Department of Physics, Faculty of Science, Ramkhamhaeng University, 282 Ramkhamhaeng Road, Bang Kapi, Bangkok 10240, Thailand \\
E-mail: danai.nuchino@hotmail.com \\
$^b$String Theory and Supergravity Group, Department
of Physics, Faculty of Science, Chulalongkorn University, 254 Phayathai Road, Pathumwan, Bangkok 10330, Thailand\\
E-mail: parinya.ka@hotmail.com \vspace{1 cm}\\
\begin{abstract}
We continue our study of gaugings the maximal $N=(2,2)$ supergravity in six dimensions with gauge groups obtained from decomposing the embedding tensor under $\mathbb{R}^+\times SO(4,4)$ subgroup of the global symmetry $SO(5,5)$. Supersymmetry requires the embedding tensor to transform in $\mathbf{144}_c$ representation of $SO(5,5)$. Under $\mathbb{R}^+\times SO(4,4)$ subgroup, this leads to the embedding tensor in $(\mathbf{8}^{\pm 3}$, $\mathbf{8}^{\pm 1},\mathbf{56}^{\pm 1})$ representations. Gaugings in $\mathbf{8}^{\pm 3}$ representations lead to a translational gauge group $\mathbb{R}^8$ while gaugings in $\mathbf{8}^{\pm 1}$ representations give rise to gauge groups related to the scaling symmetry $\mathbb{R}^+$. On the other hand, the embedding tensor in $\mathbf{56}^{\pm 1}$ representations gives $CSO(4-p,p,1)\sim SO(4-p,p)\ltimes \mathbb{R}^4\subset SO(4,4)$ gauge groups with $p=0,1,2$. More interesting gauge groups can be obtained by turning on more than one representation of the embedding tensor subject to the quadratic constraints. In particular, we consider gaugings in both $\mathbf{56}^{-1}$ and $\mathbf{8}^{+3}$ representations giving rise to larger $SO(5-p,p)$ and $SO(4-p,p+1)$ gauge groups for $p=0,1,2$. In this case, we also give a number of half-supersymmetric domain wall solutions preserving different residual symmetries. The solutions for gaugings obtained only from $\mathbf{56}^{-1}$ representation are also included in these results when the $\mathbf{8}^{+3}$ part is accordingly turned off. 
\end{abstract}
\end{center}
\newpage
\section{Introduction}
Supersymmetric domain walls are solutions to gauged supergravities that play many important roles in various aspects of string/M-theory. These solutions have provided a useful tool for studying different aspects of the AdS/CFT correspondence since the beginning of the original proposal in \cite{Maldacena}, see also \cite{subAdSCFT1,subAdSCFT2}. They are also vital in the so-called DW/QFT correspondence \cite{DWQFT1, DWQFT2, DWQFT3}, a generalization of the AdS/CFT correspondence to non-conformal field theories. In particular, these solutions give holographic descriptions to RG flows in strongly coupled dual field theories in various space-time dimensions. Domain walls also appear in the study of cosmology via the domain wall/cosmology correspondence, see for example \cite{DWcosmo1, DWcosmo2, DWcosmo3}. A systematic classification of supersymmetric domain walls from maximal gauged supergravity in various space-time dimensions has been performed in \cite{EricDW}, and many domain wall solutions in gauged supergravities have been found in different space-time dimensions, see \cite{DW1, DW2, DW3, DW4, DW5, our7DDW, our6DDW1, our6DDW2, DW6, DW7, DW8, DW9, DW10, DW11, DW12, DW13, DW14, DW15} for an incomplete list. 
\\
\indent In this paper, we are interested in domain wall solutions from maximal $N=(2,2)$ six-dimensional gauged supergravity. The ungauged $N=(2,2)$ supergravity has been constructed in \cite{Tanii}, and the first $N=(2,2)$ six-dimensional gauged supergravity with $SO(5)$ gauge group has been found in \cite{6D_Max_Cowdall} by performing an $S^1$ reduction of the maximal $SO(5)$ gauged supergravity in seven dimensions \cite{7D_Pernici}. The most general $N=(2,2)$ gauged supergravity has been constructed in \cite{6D_Max_Gauging} using the embedding tensor formalism. The embedding tensor transforms in $\mathbf{144}_c$ representation of $SO(5,5)$ global symmetry as required by supersymmetry, and some possible gaugings classified under $GL(5)$ and $SO(4,4)$ subgroups of $SO(5,5)$ have also been identified in \cite{6D_Max_Gauging}. Many gaugings arising from $GL(5)$ decomposition together with a large number of supersymmetric domain wall solutions have been constructed recently in \cite{our6DDW1} and \cite{our6DDW2}. 
\\
\indent In this  work, we will continue the study of the maximal $N=(2,2)$ gauged supergravity and the corresponding supersymmetric domain walls by considering gaugings arising from decomposing the embedding tensor under $\mathbb{R}^+\times SO(4,4)\subset SO(5,5)$. Under $\mathbb{R}^+\times SO(4,4)$, the embedding tensor in $\mathbf{144}_c$ representation of $SO(5,5)$ decomposes into $\mathbf{8}^{\pm 1}$, $\mathbf{8}^{\pm 3}$, and $\mathbf{56}^{\pm 1}$ representations of $\mathbb{R}^+\times SO(4,4)$. We will determine explicit solutions of the embedding tensor giving rise to consistent gauge groups of the $N=(2,2)$ gauged supergravity and look for possible supersymmetric domain wall solutions.  
\\
\indent According to the DW/QFT correspondence, the aforementioned domain wall solutions should be dual to maximally supersymmetric Yang-Mills theory in five dimensions. The latter plays an important role in defining the $N=(2,0)$ superconformal field theory in six dimensions compactified on $S^1$ and also describing nonperturbative dynamics of $N=1,2$, class S, theories in four dimensions, see for example \cite{SYM1,SYM2,SYM3,SYM4,SYM5,SYM6,SYM7}. We expect domain wall solutions studied here could be useful in this context as well. 
\\
\indent The paper is organized as follows. In section \ref{6DN=(2,2)gSUGRA}, we briefly review the construction of six-dimensional maximal gauged supergravity in the embedding tensor formalism. Possible gauge groups arising from decomposing the embedding tensor under $SO(4,4)$ are determined in section \ref{SO4_4_gauging}. In section \ref{DW_sol}, we find supersymmetric domain wall solutions from gaugings in $\mathbf{56}^{-1}$ and $\mathbf{8}^{+3}$ representations. Conclusions and discussions are given in section \ref{Discuss}. Relevant branching rules for $SO(5,5)$ representations under $SO(4,4)$ are given in the appendix. 
\section{$N=(2,2)$ gauged supergravity in six dimensions}\label{6DN=(2,2)gSUGRA}
We first give a brief review of six-dimensional $N=(2,2)$ gauged supergravity in the embedding tensor formalism constructed in \cite{6D_Max_Gauging}. We will only collect relevant formulae for determining possible gauge groups and finding supersymmetric domain wall solutions. For more details, we refer the reader to the original construction in \cite{6D_Max_Gauging}.
\\
\indent The supergravity multiplet of the maximal $N=(2,2)$ supersymmetry in six dimensions consists of the following component fields
\begin{equation}\label{6DSUGRAmultiplet}
\left(e^{\hat{\mu}}_\mu, B_{\mu\nu \overline{m}}, A^{A}_\mu, {V_A}^{\alpha\dot{\alpha}}, \psi_{+\mu\alpha}, \psi_{-\mu\dot{\alpha}}, \chi_{+a\dot{\alpha}}, \chi_{-\dot{a}\alpha}\right).
\end{equation}
In our convention, curved and flat space-time indices are respectively denoted by $\mu,\nu,\ldots=0,1,\ldots,5$ and $\hat{\mu},\hat{\nu},\ldots=0,1,\ldots,5$. Lower and upper $\overline{m},\overline{n},\ldots=1,\ldots,5$ indices label fundamental and anti-fundamental representations of $GL(5)\subset SO(5,5)$, respectively. Indices $A,B,\ldots =1,\ldots,16$ refer to Majorana-Weyl spinors of the $SO(5,5)$ duality symmetry. 
\\
\indent The electric two-form potentials $B_{\mu\nu \overline{m}}$, appearing in the ungauged Lagrangian, transform as $\mathbf{5}$ under $GL(5)$ while the vector fields $A^{A}_\mu$ transform as $\mathbf{16}_c$ under $SO(5,5)$. Together with the magnetic duals ${B_{\mu\nu}}^{\overline{m}}$ transforming in $\overline{\mathbf{5}}$ representation of $GL(5)$, the electric two-forms $B_{\mu\nu \overline{m}}$ transform in a vector representation $\mathbf{10}$ of the full global symmetry group $SO(5,5)$ denoted by $B_{\mu\nu M}=(B_{\mu\nu \overline{m}}, {B_{\mu\nu}}^{\overline{m}})$. Therefore, only the subgroup $GL(5)\subset SO(5,5)$ is a manifest off-shell symmetry of the theory. Indices $M,N,\ldots$ denote fundamental or vector representation of $SO(5,5)$. Finally, there are $25$ scalar fields parametrizing the coset space $SO(5,5)/SO(5)\times SO(5)$. 
\\
\indent Fermionic fields, transforming under the local $SO(5)\times SO(5)$ symmetry, are symplectic Majorana-Weyl (SMW) spinors. Indices $\alpha,\beta,\ldots$ and $\dot{\alpha},\dot{\beta},\ldots$ are respectively two sets of $SO(5)$ spinor indices in $SO(5)\times SO(5)$. Similarly, vector indices of the two $SO(5)$ factors are denoted by $a,b,\ldots$ and $\dot{a},\dot{b},\ldots$. We will use $\pm$ to indicate the space-time chiralities of the spinors. Under the local $SO(5)\times SO(5)$ symmetry, the two sets of gravitini $\psi_{+\mu\alpha}$ and $\psi_{-\mu\dot{\alpha}}$ transform as $(\mathbf{4},\mathbf{1})$ and $(\mathbf{1},\mathbf{4})$ while the spin-$\frac{1}{2}$ fields $\chi_{+a\dot{\alpha}}$ and $\chi_{-\dot{a}\alpha}$ transform as $(\mathbf{5},\mathbf{4})$ and $(\mathbf{4},\mathbf{5})$, respectively.
\\
\indent In chiral spinor representation, the $SO(5,5)/SO(5)\times SO(5)$ coset is described by a coset representative ${V_A}^{\alpha\dot{\beta}}$ transforming under the global $SO(5,5)$ and local $SO(5)\times SO(5)$ symmetries by left and right multiplications, respectively. The inverse elements ${(V^{-1})_{\alpha\dot{\beta}}}^A$ will be denoted by ${V^A}_{\alpha\dot{\beta}}$ satisfying the relations
\begin{equation}\label{VViProp}
{V_A}^{\alpha\dot{\beta}}{V^B}_{\alpha\dot{\beta}}=\delta^B_A\qquad \textrm{and}\qquad {V_A}^{\alpha\dot{\beta}}{V^A}_{\gamma\dot{\delta}}=\delta^{\alpha}_{\gamma}\delta^{\dot{\beta}}_{\dot{\delta}}\, .
\end{equation}
On the other hand, in vector representation, the coset representative is given by a $10\times10$ matrix ${\mathcal{V}_M}^{\underline{A}}=({\mathcal{V}_M}^{a}, {\mathcal{V}_M}^{\dot{a}})$ with $\underline{A}=(a,\dot{a})$ and related to the coset representative in chiral spinor representation by the following relations
\begin{eqnarray}
{\mathcal{V}_M}^a&=&\frac{1}{16}V^{A\alpha\dot{\alpha}}(\Gamma_M)_{AB}{(\gamma^a)_{\alpha\dot{\alpha}}}^{\beta\dot{\beta}}{V^B}_{\beta\dot{\beta}},\label{VVrel1}\\{\mathcal{V}_M}^{\dot{a}}&=&-\frac{1}{16}V^{A\alpha\dot{\alpha}}(\Gamma_M)_{AB}{(\gamma^{\dot{a}})_{\alpha\dot{\alpha}}}^{\beta\dot{\beta}}{V^B}_{\beta\dot{\beta}}\, .\label{VVrel2}
\end{eqnarray}
In these equations, $(\Gamma_M)_{AB}$ and ${(\Gamma_{\ul{A}})_{\alpha\dot{\alpha}}}^{\beta\dot{\beta}}=({(\gamma_a)_{\alpha\dot{\alpha}}}^{\beta\dot{\beta}},{(\gamma_{\dot{a}})_{\alpha\dot{\alpha}}}^{\beta\dot{\beta}})$ are respectively $SO(5,5)$ gamma matrices in non-diagonal $\eta_{MN}$ and diagonal $\eta_{\ul{A}\ul{B}}$ bases. The inverse of ${\mathcal{V}_M}^{\underline{A}}$ will be denoted by $\mathcal{V}^{M\underline{A}}$ satisfying the following relations 
\begin{equation}
\mathcal{V}^{Ma}{\mathcal{V}_M}^b=\delta^{ab},\qquad\mathcal{V}^{M\dot{a}}{\mathcal{V}_M}^{\dot{b}}=\delta^{\dot{a}\dot{b}},\qquad\mathcal{V}^{Ma}{\mathcal{V}_M}^{\dot{a}}=0
\end{equation}
and
\begin{equation}
{\mathcal{V}_M}^a\mathcal{V}^{Na}-{\mathcal{V}_M}^{\dot{a}}\mathcal{V}^{N\dot{a}}=\delta^N_M\, .\label{V_Vi_vec}
\end{equation}
In these equations, we have explicitly raised the $SO(5)\times SO(5)$ vector index $\underline{A}=(a,\dot{a})$ resulting in a minus sign in equation \eqref{V_Vi_vec}.
\\
\indent The most general gaugings of six-dimensional $N=(2,2)$ supergravity are described by the embedding tensor in $\mathbf{144}_c$ representation of $SO(5,5)$. This can be written in terms of a vector-spinor of $SO(5,5)$, $\theta^{AM}$, subject to the linear constraint (LC)
\begin{equation}\label{MainLC}
(\Gamma_M)_{AB}\,\theta^{BM}\ =\ 0\,.
\end{equation}
The gauge covariant derivative is then given by
\begin{equation}\label{gauge_covariant_derivative}
D_\mu=\partial_\mu-gA^A_\mu\ {\Theta_A}^{MN}\boldsymbol{t}_{MN}
\end{equation}
with $g$ being a gauge coupling constant and ${\Theta_A}^{MN}$ defined by 
\begin{equation}
{\Theta_A}^{MN}\ =\ -\theta^{B[M}(\Gamma^{N]})_{BA}\ \equiv \ \left(\Gamma^{[M}\theta^{N]}\right)_A\, .
\end{equation}
\indent As usual, the embedding tensor identifies generators $X_A={\Theta_A}^{MN}\boldsymbol{t}_{MN}$ of the gauge group $G_0\subset SO(5,5)$ with particular linear combinations of the $SO(5,5)$ generators $\boldsymbol{t}_{MN}$. Consistency also requires the gauge generators to form a closed subalgebra of $SO(5,5)$ implying the quadratic constraint (QC)
\begin{equation}
\left[X_A,X_B\right]\ = \ -{(X_A)_B}^C\,X_C\, .
\end{equation} 
In terms of $\theta^{AM}$, this constraint reduces to the following two conditions
\begin{eqnarray}
\theta^{AM}\theta^{BN}\eta_{MN}&=&0,\label{QC1}\\ 
\theta^{AM}\theta^{B[N}(\Gamma^{P]})_{AB}&=&0\, .\label{QC2}
\end{eqnarray}  
Any $\theta^{AM}\in\mathbf{144}_c$ satisfying these quadratic constraints defines a consistent gauging.
\\
\indent In this work, we are only interested in classifying possible gauge groups and finding supersymmetric domain wall solutions involving only the metric and scalar fields. We have explicitly checked that the truncation of vector and tensor fields is consistent in all the domain wall solutions given in section \ref{DW_sol}. This follows from the fact that the corresponding Yang-Mills currents vanish for all of the solutions considered here. With all vector and tensor fields set to zero, the bosonic Lagrangian of the maximal $N=(2,2)$ gauged supergravity is given by
\begin{equation}
e^{-1}\mathcal{L}=\frac{1}{4}R-\frac{1}{16}P_{\mu}^{a\dot{a}}P^\mu_{a\dot{a}}-\mathbf{V}\label{bosonic_L}
\end{equation}
while the supersymmetry transformations of fermionic fields read
\begin{eqnarray}
\delta\psi_{+\mu\alpha}&=& D_\mu\epsilon_{+\alpha}+\frac{g}{4}\hat{\gamma}_\mu {T_\alpha}^{\dot{\beta}}\epsilon_{-\dot{\beta}},\label{1stSUSY}\\
\delta\psi_{-\mu\dot{\alpha}}&=& D_\mu\epsilon_{-\dot{\alpha}}-\frac{g}{4}\hat{\gamma}_\mu {T^{\beta}}_{\dot{\alpha}}\epsilon_{+\beta},\label{2ndSUSY}\\
\delta\chi_{+a\dot{\alpha}}&=&\frac{1}{4}P^\mu_{a\dot{a}}\hat{\gamma}_\mu{(\gamma^{\dot{a}})_{\dot{\alpha}}}^{\dot{\beta}}\epsilon_{-\dot{\beta}}+2g{(T_{a})^\beta}_{\dot{\alpha}}\epsilon_{+\beta}-\frac{g}{2}{T^{\alpha}}_{\dot{\alpha}}{(\gamma_a)_\alpha}^\beta\epsilon_{+\beta},\label{3rdSUSY}\\
\delta\chi_{-\dot{a}\alpha}&=&\frac{1}{4}P^\mu_{a\dot{a}}\hat{\gamma}_\mu{(\gamma^a)_\alpha}^\beta\epsilon_{+\beta}+2g{(T_{\dot{a}})_{\alpha}}^{\dot{\beta}}\epsilon_{-\dot{\beta}}+\frac{g}{2}{T_{\alpha}}^{\dot{\alpha}}{(\gamma_{\dot{a}})_{\dot{\alpha}}}^{\dot{\beta}}\epsilon_{-\dot{\beta}}.\label{4thSUSY}
\end{eqnarray}
The covariant derivatives of supersymmetry parameters are defined as
\begin{eqnarray}
D_\mu\epsilon_{+\alpha}&=& \partial_\mu\epsilon_{+\alpha}+\frac{1}{4}{\omega_\mu}^{\nu\rho}\hat{\gamma}_{\nu\rho}\epsilon_{+\alpha}+\frac{1}{4}Q_\mu^{ab}{(\gamma_{ab})_\alpha}^\beta\epsilon_{+\beta},\label{CoDivEp+}\\
D_\mu\epsilon_{-\dot{\alpha}}&=& \partial_\mu\epsilon_{-\dot{\alpha}}+\frac{1}{4}{\omega_\mu}^{\nu\rho}\hat{\gamma}_{\nu\rho}\epsilon_{-\dot{\alpha}}+\frac{1}{4}Q_\mu^{\dot{a}\dot{b}}{(\gamma_{\dot{a}\dot{b}})_{\dot{\alpha}}}^{\dot{\beta}}\epsilon_{-\dot{\beta}}\label{CoDivEp-}
\end{eqnarray}
with $\hat{\gamma}_\mu=e_\mu^{\hat{\mu}}\hat{\gamma}_{\hat{\mu}}$. $\hat{\gamma}_{\hat{\mu}}$ are space-time gamma matrices, and for simplicity, we will suppress all space-time spinor indices.
\\
\indent The scalar vielbein $P_{\mu}^{a\dot{a}}$ and $SO(5)\times SO(5)$ composite connections, $Q_\mu^{ab}$ and $Q_\mu^{\dot{a}\dot{b}}$, are given by
\begin{eqnarray}
P_{\mu}^{a\dot{a}}&=&\frac{1}{4}{(\gamma^a)}^{\alpha\beta}{(\gamma^{\dot{a}})}^{\dot{\alpha}\dot{\beta}}{V^A}_{\alpha\dot{\alpha}}\partial_\mu V_{A\beta\dot{\beta}},\label{PDef}\\
Q_{\mu}^{ab}&=&\frac{1}{8}{(\gamma^{ab})}^{\alpha\beta}\Omega^{\dot{\alpha}\dot{\beta}} {V^A}_{\alpha\dot{\alpha}}\partial_\mu V_{A\beta\dot{\beta}},\label{QuDef}\\
Q_{\mu}^{\dot{a}\dot{b}}&=&\frac{1}{8}\Omega^{\alpha\beta}{(\gamma^{\dot{a}\dot{b}})}^{\dot{\alpha}\dot{\beta}}{V^A}_{\alpha\dot{\alpha}}\partial_\mu V_{A\beta\dot{\beta}}\, .\label{QdDef}
\end{eqnarray}
In these equations, $\Omega^{\alpha\beta}$ and $\Omega^{\dot{\alpha}\dot{\beta}}$ are the $USp(4)\sim SO(5)$ symplectic forms that satisfy the following relations
\begin{equation}
\Omega_{\beta\alpha}\ =\ -\Omega_{\alpha\beta},\qquad \Omega^{\alpha\beta}\ =\ (\Omega_{\alpha\beta})^*,\qquad \Omega_{\alpha\beta}\Omega^{\beta\gamma}\ =\ -\delta^\gamma_\alpha
\end{equation}
and similarly for $\Omega_{\dot{\alpha}\dot{\beta}}$. The scalar potential is given by
\begin{eqnarray}\label{scalarPot}
\mathbf{V}&=&\frac{g^2}{2}\theta^{AM}\theta^{BN}{\mathcal{V}_M}^a{\mathcal{V}_N}^b\left[{V_A}^{\alpha\dot{\alpha}}{(\gamma_a)_\alpha}^\beta{(\gamma_b)_\beta}^\gamma V_{B\gamma\dot{\alpha}}\right]\nonumber\\
&=&-\frac{g^2}{2}\left[T^{\alpha\dot{\alpha}}T_{\alpha\dot{\alpha}}-2(T^a)^{\alpha\dot{\alpha}}(T_a)_{\alpha\dot{\alpha}}\right]\label{scalarPot}
\end{eqnarray}
with the T-tensors defined by
\begin{equation}\label{TTenDef}
(T^a)^{\alpha\dot{\alpha}}={\mathcal{V}_M}^a\theta^{AM}{V_A}^{\alpha\dot{\alpha}},\qquad (T^{\dot{a}})^{\alpha\dot{\alpha}}=-{\mathcal{V}_M}^{\dot{a}}\theta^{AM}{V_A}^{\alpha\dot{\alpha}},
\end{equation}
and
\begin{equation}
T^{\alpha\dot{\alpha}}\equiv (T^a)^{\beta\dot{\alpha}}{(\gamma_a)_\beta}^\alpha=-(T^{\dot{a}})^{\alpha\dot{\beta}}{(\gamma_{\dot{a}})_{\dot{\beta}}}^{\dot{\alpha}}\, .
\end{equation}
We also note useful identities involving various components of the T-tensors
\begin{eqnarray}
& &D_\mu T^a=\frac{1}{4}P^{b\dot{b}}_\mu \left(\gamma^bT^a\gamma^{\dot{b}}-2\delta^{ab}T^{\dot{b}}\right),\\
& &D_\mu T^{\dot{a}}=\frac{1}{4}P^{b\dot{b}}_\mu \left(\gamma^bT^{\dot{a}}\gamma^{\dot{b}}-2\delta^{\dot{a}\dot{b}}T^{b}\right),\\
& &D_\mu T=\frac{1}{2}P^{a\dot{a}}_\mu \left(T^a\gamma^{\dot{a}}-\gamma^aT^{\dot{a}}-\frac{1}{2}\gamma^aT\gamma^{\dot{a}}\right). \label{T_iden3}
\end{eqnarray}

\section{Gaugings of six-dimensional $N=(2,2)$ supergravity under $SO(4,4)$}\label{SO4_4_gauging}
In this section, we will determine explicit forms of the embedding tensor for a number of possible gauge groups leading to consistent $N=(2,2)$ gauged supergravities in six dimensions. To find gauge groups by decomposing the embedding tensor under $\mathbb{R}^+\times SO(4,4)\subset SO(5,5)$, we decompose the $SO(5,5)$ vector index as $M=(-,I,+)$ with $I=1,2,\ldots, 8$ being the $SO(4,4)$ vector index. The $SO(5,5)$ generators $\boldsymbol{t}_{MN}$ are decomposed accordingly as $\boldsymbol{t}_{MN}=(\boldsymbol{t}_{-+}=\boldsymbol{d},\boldsymbol{t}_{+I}=\boldsymbol{p}_{I},\boldsymbol{t}_{-I}=\boldsymbol{k}_{I},\boldsymbol{t}_{IJ}=\boldsymbol{\tau}_{IJ})$ with $\boldsymbol{d}$ and $\boldsymbol{\tau}_{IJ}$ being $\mathbb{R}^+$ and $SO(4,4)$ generators, respectively.
\\
\indent Similarly, the $SO(5,5)$ spinor index $A$ will also be split as $A=(m,\dot{m})$ with $m=1,2,\ldots, 8$ and $\dot{m}=\dot{1},\dot{2},\ldots, \dot{8}$ being $SO(4,4)$ spinor indices, see more detail in the appendix. As given in the appendix, the embedding tensor transforming in $\mathbf{144}_c$ representation of $SO(5,5)$ will split into the following representations under $\mathbb{R}^+\times SO(4,4)$ 
\begin{equation}
\underbrace{\mathbf{144}_c}_{\theta^{AM}}\ \rightarrow\ \underbrace{\mathbf{56}^{-1}}_{\vartheta_3^{m I}}\,\oplus\,\underbrace{\mathbf{56}^{+1}}_{\vartheta_4^{\dot{m} I}}\,\oplus\,\underbrace{\mathbf{8}^{-1}}_{\theta_6^{\dot{m} +}}\,\oplus\,\underbrace{\mathbf{8}^{+1}}_{\theta_1^{m -}}\,\oplus\,\underbrace{\mathbf{8}^{+3}}_{\theta_2^{\dot{m} -}}\,\oplus\,\underbrace{\mathbf{8}^{-3}}_{\theta_5^{m +}}.
\end{equation}
For convenience, we also recall the identification of various components of the embedding tensor of the form
\begin{equation}
\theta^{AM}\ = \ \left(\begin{array}{c|c|c} \theta_1^{m -}  & \theta_3^{m I} & \theta_5^{m +} \\\hline
								\theta_2^{\dot{m} -} & \theta_4^{\dot{m} I} & \theta_6^{\dot{m} +\phantom{'}} \end{array}\right).
\end{equation}
The components $\theta_1^{m -}$, $\theta_4^{\dot{m} I}$, and $\theta_5^{m +}$ correspond to $\mathbf{8}^{+1}$, $\mathbf{56}^{+1}$, and $\mathbf{8}^{-3}$ representations while $\theta_2^{\dot{m} -}$, $\theta_3^{m I}$, and $\theta_6^{\dot{m} +}$ are respectively $\mathbf{8}^{+3}$, $\mathbf{56}^{-1}$, and $\mathbf{8}^{-1}$ ones. The linear combinations $\vartheta_3^{m I}$ and ${\vartheta_4^{\dot{m} I}}$ in terms of $(\theta^{mI}_3,\theta^{\dot{m}+}_6)$ and $(\theta^{n-}_1,\theta^{\dot{m}I}_4,)$, as required by the LC, are defined in \eqref{vartheta_def}. For later convenience, we also repeat these relations here
\begin{equation}
\vartheta_3^{m I}=\theta_3^{m I}+\frac{\sqrt{2}}{8}{(\gamma^I)^{m}}_{\dot{n}}\theta_6^{\dot{n} +}\qquad\text{and }\qquad\vartheta_4^{\dot{m} I}=\theta_4^{\dot{m} I}-\frac{\sqrt{2}}{8}{(\gamma^I)^{\dot{m}}}_n\theta_1^{n -}\, .
\end{equation} 
\indent In this section, we will determine explicit forms of the embedding tensor by imposing the quadratic constraint on the embedding tensor. With the above decomposition, the first condition of QC given in \eqref{QC1} reduces to
\begin{equation}\label{QC1a}
\theta^{A+}\theta^{B-}+\theta^{AI}\theta^{BJ}\eta_{IJ}+\theta^{A-}\theta^{B+}=0
\end{equation}
with $\eta_{IJ}$ being the $SO(4,4)$ invariant tensor defined in \eqref{SO(4,4)off-diag-eta}. On the other hand, the second condition \eqref{QC2} splits into
\begin{eqnarray}
\theta^{m M}\theta_1^{n -}\mathbf{c}_{mn}+\theta^{\dot{m} M}\theta_6^{\dot{n} +}\mathbf{c}_{\dot{m}\dot{n}}&=&0,\label{QC2a}\\
\theta^{m M}\theta_2^{\dot{n} -}(\gamma^I)_{m\dot{n}}+\theta^{\dot{m} M}\theta_1^{n -}(\gamma^I)_{\dot{m}n}-\sqrt{2}\theta^{\dot{m} M}\theta_4^{\dot{n} I}\mathbf{c}_{\dot{m}\dot{n}}&=&0,\label{QC2b}\\
\theta^{m M}\theta_6^{\dot{n} +}(\gamma^I)_{m\dot{n}}+\theta^{\dot{m} M}\theta_5^{n +}(\gamma^I)_{\dot{m}n}+\sqrt{2}\theta^{m M}\theta_3^{n I}\mathbf{c}_{mn}&=&0,\label{QC2c}\\
\theta^{m M}\theta_4^{\dot{n} [I}(\gamma^{J]})_{m\dot{n}}+\theta^{\dot{m} M}\theta_3^{n [I}(\gamma^{J]})_{\dot{m}n}&=&0\, .\label{QC2d}
\end{eqnarray}
In these equations, $\mathbf{c}_{mn}$ and $\mathbf{c}_{\dot{m}\dot{n}}$ are elements of the $SO(4,4)$ charge conjugation matrix defined in \eqref{SO4_4_CC}, and $(\gamma^I)_{m\dot{n}}=(\gamma^I)_{\dot{n}m}$ are chirally decomposed $SO(4,4)$ gamma matrices given in \eqref{gammaIdecom2} 
\\
\indent Some possible gauge groups under $SO(4,4)$ have also been discussed in \cite{6D_11}, and it has been pointed out that turning on only $\theta_1^{m -}$ or $\theta_6^{\dot{m} +}$ components leads to gaugings of the scaling symmetry $\mathbb{R}^+$. Furthermore, with $\theta_4^{\dot{m} I}=0$ or $\theta_3^{m I}=0$, we find from the LC given in \eqref{red_LC} that $\theta_1^{m -}$ or $\theta_6^{\dot{m} +}$ need to be zero, respectively. Accordingly, we conclude that gaugings with only $\mathbf{8}^{+1}$ or $\mathbf{8}^{-1}$ components non-vanishing are not consistent.
 
\subsection{Gaugings in $\mathbf{8}^{+3}$ representation}\label{8+3gauging}
We begin with gauge groups arising from the embedding tensor in $\mathbf{8}^{+3}$ representation. In this case, we set all $\theta$'s components to be zero except
\begin{equation}
\theta_2^{\dot{m} -}=v^{\dot{m}}
\end{equation}
for a spinor $v^{\dot{m}}$. The $\theta^{AM}$ matrix of the form
\begin{equation}
\theta^{AM}\ = \ \left(\begin{array}{c|c|c}   & \phantom{\theta_3^{\alpha I}} &  \\\hline
								v^{\dot{m}} &  & \phantom{\theta_6^{\dot{\alpha}}} \end{array}\right)
\end{equation}
makes the embedding tensor satisfy all the LC and QC. We have used the notation that all vanishing elements are left as blank spaces. 
\\
\indent For $A=(m,\dot{m})$, the corresponding gauge generators split into $X_A=(X_m,\,X_{\dot{m}})$. With the above embedding tensor, the last eight generators vanish, $X_{\dot{m}}=0$, while the first eight generators are given in terms of $\boldsymbol{k}_{I}$ as
\begin{equation}\label{first8Xm}
X_m=(\gamma^I)_{m\dot{n}}v^{\dot{n}}\boldsymbol{k}_{I}\, .
\end{equation}
They are all linearly independent and commute with each other $\left[ X_m,X_n\right] =0$. Thus, the resulting gauge group is an eight-dimensional translational group $\mathbb{R}^8$ associated with the $\boldsymbol{k}_{I}$ generators.

\subsection{Gaugings in $\mathbf{8}^{-3}$ representation}
As in the previous case, we set all $\theta$'s components to be zero except
\begin{equation}
\theta_5^{m +}=w^{m}
\end{equation}
for any spinor $w^{m}$. All the LC and QC are satisfied by this embedding tensor. In this case, there are also eight non-vanishing gauge generators, but given in terms of the $\boldsymbol{p}_{I}$ generators, i.e.
\begin{equation}
X_{\dot{m}}=(\gamma^I)_{\dot{m}n}w^{n}\boldsymbol{p}_{I}\, .
\end{equation}
As in the previous case, they are all linearly independent and commute with each other, $\left[ X_{\dot{m}},X_{\dot{n}}\right] =0$. This implies again that the resulting gauge group is an eight-dimensional translational group $\mathbb{R}^8$ associated with the $\boldsymbol{p}_{I}$ generators.

\subsection{Gaugings in $\mathbf{56}^{-1}$ representation}\label{Sec56min}
We now move to gaugings in $\mathbf{56}^{-1}$ representation by choosing only $\theta_3^{m I}$ to be non-vanishing. The embedding tensor takes the form
\begin{equation}
\theta^{AM}\ = \ \left(\begin{array}{c|c|c}   & \theta_3^{m I} & \phantom{\theta_6^{\dot{\alpha}}} \\\hline
								\phantom{\theta_3^{\alpha I}} &  &  \end{array}\right)
\end{equation}
subject to the LC 
\begin{equation}\label{56m_LC}
(\gamma_I)_{\dot{m}n}\theta_3^{n I}=0
\end{equation}
which is the same as the second condition in \eqref{var_red_LC}. For vanishing $\theta_6^{\dot{m} +}$ component from $\mathbf{8}^{-1}$ representation, the embedding tensor is simply given by $\theta_3^{m I}$. As in \cite{6D_11}, all $56$ components in $\theta_3^{m I}$ can be parametrized by an antisymmetric tensor $f_{\dot{m}\dot{n}\dot{p}}=f_{[\dot{m}\dot{n}\dot{p}]}$ by writing $\theta_3^{m I}$ as
\begin{equation}\label{theta3PARA}
\theta_3^{m I}=\frac{1}{48}f_{\dot{m}\dot{n}\dot{p}}(\gamma^{IJ})^{\dot{m}\dot{n}}(\gamma_J)^{m\dot{p}}
\end{equation}
with $(\gamma^{IJ})^{\dot{m}\dot{n}}=(\gamma^{[I})^{\dot{m}p}{(\gamma^{J]})_p}^{\dot{n}}$. 
\\
\indent The LC given in \eqref{56m_LC} is now identically satisfied, and the corresponding gauge generators are split into the following two sets
\begin{eqnarray}
X_m&=&\frac{1}{24\sqrt{2}}f_{\dot{m}\dot{n}\dot{p}}(\gamma^{IJ})^{\dot{m}\dot{n}}{(\gamma_J)_m}^{\dot{p}}\boldsymbol{p}_{I},\label{CSO(4,0,1)GG1}\\
X_{\dot{m}}&=&\frac{1}{48}f_{\dot{n}\dot{p}\dot{q}}(\gamma^{IK})^{\dot{n}\dot{p}}(\gamma_K)^{q\dot{q}}(\gamma^J)_{q\dot{m}}\boldsymbol{\tau}_{IJ}.\label{CSO(4,0,1)GG2}
\end{eqnarray}
The first set contains eight nilpotent generators that commute with each other, $\left[X_m,\,X_n\right]=0$, so they generate a translational subgroup associated with $\boldsymbol{p}_{I}$ generators. The other set gives another subgroup embedded in the $SO(4,4)$ factor. According to \cite{6D_11}, the QCs in terms of the antisymmetric tensor $f_{\dot{m}\dot{n}\dot{p}}$ can be written as
\begin{equation}\label{RedQC}
f_{\dot{m}\dot{n}\dot{p}}f^{\dot{m}\dot{n}\dot{p}}=0\qquad\text{ and }\qquad f_{\dot{r}[\dot{m}\dot{n}}{f_{\dot{p}\dot{q}]}}^{\dot{r}}=0\, .
\end{equation}
We will discuss some possible solutions to these conditions.

\subsubsection{$CSO(4,0,1)\sim SO(4)\ltimes\mathbb{R}^4$ gauge group}\label{CSO401Sec}
We first consider a simple solution of the form
\begin{equation}
f_{\dot{m}\dot{n}\dot{p}}=(\kappa_1\varepsilon_{ijk},\kappa_2\varepsilon_{rst})
\end{equation}
for $i,j,\ldots=\dot{1},\dot{2},\dot{3}$ and $r,s,\ldots=\dot{5},\dot{6},\dot{7}$. To solve the first condition in \eqref{RedQC}, we need to impose the relation $\kappa_1=\pm \kappa_2$. We will choose $\kappa_1=\kappa_2=\kappa\in \mathbb{R}$ for definiteness.
\\
\indent With this form of the embedding tensor, we find that the gauge generators $X_{\dot{4}}$ and $X_{\dot{8}}$ vanish. Commutation relations between $X_i$ and $X_r$ lead directly to $SO(3)\times SO(3)$ algebra
\begin{equation}
\left[X_i,\,X_j\right]=-\kappa\varepsilon_{ijk}X_k,\qquad\left[X_r,\,X_s\right]=\kappa\varepsilon_{rst}X_t,\qquad\left[X_i,\,X_r\right]=0\, .
\end{equation}
The remaining eight generators correspond to translational generators, but in this case, there are four constraints among them
\begin{equation}\label{irrCSO401TranCon}
X_5=X_1,\qquad X_6=-X_2,\qquad X_7=-X_3,\qquad X_8=-X_4\, .
\end{equation}
Therefore, there are only four linearly independent translational generators. 
\\
\indent To make the form of the resulting gauge group explicit, we redefine the gauge generators as follows. We first introduce the $SO(4)\sim SO(3)\times SO(3)$ generators $M_{\tilde{\mu}\tilde{\nu}}=-M_{\tilde{\nu}\tilde{\mu}}$ for $\tilde{\mu},\tilde{\nu}=1,2,3,4$. These satisfy the standard $SO(4)$ algebra of the form
\begin{equation}\label{simSO(4)algebra}
\left[M_{\tilde{\mu}\tilde{\nu}},M_{\tilde{\rho}\tilde{\sigma}}\right]=2\kappa\left(\delta_{\tilde{\mu}[\tilde{\rho}}M_{\tilde{\sigma}]\tilde{\nu}}-\delta_{\tilde{\nu}[\tilde{\rho}}M_{\tilde{\sigma}]\tilde{\mu}}\right).
\end{equation}
In terms of $X_i$ and $X_r$ generators, we find, for $\tilde{\mu}=(i,4)$,
\begin{equation}\label{SO4M3}
M_{ij}=\varepsilon_{ijk}A_k\qquad \textrm{and}\qquad M_{4i}=-M_{i4}=B_i
\end{equation}
with
\begin{eqnarray}\label{irrSO(4)gen}
A_1&\hspace{-0.1cm}=&\hspace{-0.1cm}X_{\dot{1}}-X_{\dot{5}},\qquad A_2=X_{\dot{2}}-X_{\dot{6}},\qquad A_3=X_{\dot{3}}-X_{\dot{7}},\nonumber\\
B_1&\hspace{-0.1cm}=&\hspace{-0.1cm}X_{\dot{1}}+X_{\dot{5}},\qquad B_2=X_{\dot{2}}+X_{\dot{6}},\qquad B_3=X_{\dot{3}}+X_{\dot{7}}\, .
\end{eqnarray}
Furthermore, we redefine the four independent translational generators as
\begin{equation}\label{3rankindytrans}
K_1=X_4,\qquad K_2=-X_3,\qquad K_3=X_2,\qquad K_4=X_1
\end{equation}
and obtain the following commutation relations 
\begin{equation}\label{Tran_SO(4)_com}
\left[K_{\tilde{\mu}},M_{\tilde{\nu}\tilde{\rho}}\right]=2\kappa\delta_{\tilde{\mu}[\tilde{\nu}}\delta_{\tilde{\rho}]}^{\tilde{\sigma}}K_{\tilde{\sigma}}\, .
\end{equation}
This implies that the gauge group takes the form of
\begin{equation}
SO(4)\ltimes\mathbb{R}^4\sim CSO(4,0,1).
\end{equation} 

\subsubsection{$CSO(3,1,1)\sim SO(3,1)\ltimes\mathbb{R}^4$ gauge group}\label{CSO311Sec}
There is another solution to the conditions \eqref{RedQC} with the antisymmetric tensor $f_{\dot{m}\dot{n}\dot{p}}$ of the form
\begin{equation}
f_{\dot{m}\dot{n}\dot{p}}=\kappa(\varepsilon_{ijr},\varepsilon_{irs})
\end{equation}
for $i,j,\ldots=\dot{1},\dot{2},\dot{3}$ and $r,s,\ldots=\dot{5},\dot{6},\dot{7}$. As in the previous case, there are eight nilpotent generators $X_m$ subject to four constraints given in \eqref{irrCSO401TranCon} together with six non-vanishing gauge generators $X_i$ and $X_r$. The latter satisfy the following commutation relations
 \begin{equation}
\left[X_i,\,X_j\right]=\kappa\varepsilon_{ijr}X_r,\quad\left[X_r,\,X_s\right]=-\kappa\varepsilon_{rsi}X_i,\quad\
\left[X_i,\,X_r\right]=\kappa\varepsilon_{irk}(\delta_{ks}X_s-X_k)
\end{equation}
with $\delta_{ir}=\text{diag}(1,1,1)$. These relations correspond to an $SO(3,1)$ algebra which can be explicitly seen by defining
\begin{equation}
M_{ij}=\varepsilon_{ijk}(\delta_{kr}X_r-X_k)\qquad\textrm{and} \qquad M_{4i}=-M_{i4}=-\frac{1}{\sqrt{3}}(\delta_{ir}X_r+X_i).
\end{equation}
These generators satisfy the $SO(3,1)$ algebra
\begin{equation}\label{simSO(3,1)algebra}
\left[M_{\tilde{\mu}\tilde{\nu}},M_{\tilde{\rho}\tilde{\sigma}}\right]=2\kappa\left(\eta_{\tilde{\mu}[\tilde{\rho}}M_{\tilde{\sigma}]\tilde{\nu}}-\eta_{\tilde{\nu}[\tilde{\rho}}M_{\tilde{\sigma}]\tilde{\mu}}\right)
\end{equation}
with $\eta_{\tilde{\mu}\tilde{\nu}}=\text{diag}(1,1,1,-1)$ for $\tilde{\mu},\tilde{\nu}=1,2,3,4$.
\\
\indent Redefining the four linearly independent translational generators as 
\begin{equation}\label{SO31K}
K_1=X_4,\qquad K_2=-X_3,\qquad K_3=X_2,\qquad K_4=\frac{X_1}{\sqrt{3}},
\end{equation}
we find the commutation relations with the $SO(3,1)$ generators 
\begin{equation}\label{Tran_SO(3,1)_com}
\left[K_{\tilde{\mu}},M_{\tilde{\nu}\tilde{\rho}}\right]=2\kappa\eta_{\tilde{\mu}[\tilde{\nu}}\delta_{\tilde{\rho}]}^{\tilde{\sigma}}K_{\tilde{\sigma}}\, .
\end{equation}
Accordingly, the resulting gauge group is given by
\begin{equation}
SO(3,1)\ltimes\mathbb{R}^4\sim CSO(3,1,1).
\end{equation} 

\subsubsection{$CSO(2,2,1)\sim SO(2,2)\ltimes\mathbb{R}^4$ gauge group}\label{CSO221Sec}
As a final example for solutions to \eqref{RedQC}, we consider the embedding tensor of the form
\begin{equation}
f_{\dot{m}\dot{n}\dot{p}}=\kappa(\varepsilon_{\bar{i}\bar{j}\bar{k}},\varepsilon_{\bar{r}\bar{s}\bar{t}})
\end{equation}
with $\bar{i},\bar{j},...=\dot{1},\dot{6},\dot{7}$ and $\bar{r},\bar{s},...=\dot{2},\dot{3},\dot{5}$ being two sets of indices that can be raised and lowered by $\eta^{\bar{i}\bar{j}}=\eta_{\bar{i}\bar{j}}=\text{diag}(-1,1,1)$ and $\eta^{\bar{r}\bar{s}}=\eta_{\bar{r}\bar{s}}=\text{diag}(1,1,-1)$, respectively. There are again eight nilpotent generators $X_m$ subject to four constraints given in \eqref{irrCSO401TranCon}. We will choose the independent generators to be 
\begin{equation}\label{SO22K}
K_1=-X_3,\qquad K_2=X_2,\qquad K_3=X_4,\qquad K_4=X_1\, .
\end{equation}
Commutation relations between the remaining six non-vanishing gauge generators $X_{\bar{i}}$ and $X_{\bar{r}}$ are given by 
\begin{equation}
\left[X_{\bar{i}},\,X_{\bar{j}}\right]=\kappa\varepsilon_{\bar{i}\bar{j}\bar{k}}\eta^{\bar{k}\bar{l}}X_{\bar{l}},\qquad\left[X_{\bar{r}},\,X_{\bar{s}}\right]=\kappa\varepsilon_{\bar{r}\bar{s}\bar{t}}\eta^{\bar{t}\bar{u}}X_{\bar{u}},\qquad\left[X_{\bar{i}},\,X_{\bar{r}}\right]=0\, .
\end{equation}
These lead to $SO(2,1)\times SO(2,1)\sim SO(2,2)$ algebra. As in the previous cases, we can also redefine the generators as
\begin{equation}\label{CSO221M}
M_{i4}=-M_{4i}=\delta_{ij}G^j+H_i\qquad\text{ and } \qquad M_{ij}=\varepsilon_{ijk}\eta^{kl}(\delta_{lm}G^m-H_l)
\end{equation}
for $i,j,\ldots=1,2,3$ and $\eta^{ij}=\eta_{ij}=\text{diag}(1,1,-1)$ together with
\begin{eqnarray}
& &G^1=X_{\dot{6}},\qquad G^2=X_{\dot{7}},\qquad G^3=X_{\dot{1}},\nonumber \\
& & H_1=X_{\dot{2}},\qquad H_2=X_{\dot{3}},\qquad H_3=X_{\dot{5}}\, .\label{CSO221GH}
\end{eqnarray}
These generators satisfy the algebra of the form given in \eqref{simSO(3,1)algebra} but with $\eta_{\tilde{\mu}\tilde{\nu}}=\text{diag}(1,1,-1,-1)$. 
\\
\indent Finally, the commutation relations between these $SO(2,2)$ generators and the four nilpotent generators given in \eqref{SO22K} are the same as in \eqref{Tran_SO(3,1)_com} for $\eta_{\tilde{\mu}\tilde{\nu}}=\text{diag}(1,1,-1,-1)$. Consequently, the resulting gauge group is given by
\begin{equation}
SO(2,2)\ltimes\mathbb{R}^4\sim CSO(2,2,1).
\end{equation} 
In summary, the embedding tensor in $\mathbf{56}^{-1}$ representation can lead to $CSO(4-p,p,1)\sim SO(4-p,p)\ltimes \mathbb{R}^4$ gauge group for $p=0,1,2$.

\subsection{Gaugings in $\mathbf{56}^{+1}$ representation}\label{App56p}
Similar to the previous case, gaugings in $\mathbf{56}^{+1}$ representation can be obtained by turning on only $\theta_4^{\dot{m} I}$ component with the same parametrization as in \eqref{theta3PARA}
\begin{equation}\label{theta4PARA}
\theta_4^{\dot{m} I}=\frac{1}{48}f_{mnp}(\gamma^{IJ})^{mn}(\gamma_J)^{\dot{m}p}
\end{equation}
where $(\gamma^{IJ})^{mn}=(\gamma^{[I})^{m\dot{p}}{(\gamma^{J]})_{\dot{p}}}^n$. The LC requires $f_{mnp}=f_{[mnp]}$ and the QCs reduce to
\begin{equation}
f_{mnp}f^{mnp}=0\qquad\text{ and }\qquad f_{r[mn}{f_{pq]}}^{r}=0.
\end{equation}
We can repeat the same analysis as in the case of gaugings from $\mathbf{56}^{-1}$ representation by solving these QCs and find the same $CSO(4-p,p,1)\sim SO(4-p,p)\ltimes \mathbb{R}^4$ gauge group for $p=0,1,2$. However, the gauge generators for the two sets of nilpotent and $SO(4-p,p)$ generators are interchanged as
\begin{eqnarray}
X_m&=&\frac{1}{48}f_{npq}(\gamma^{IK})^{np}(\gamma_K)^{\dot{q}q}(\gamma^J)_{m\dot{q}}\boldsymbol{\tau}_{IJ},\\
X_{\dot{m}}&=&\frac{1}{24\sqrt{2}}f_{mnp}(\gamma^{IJ})^{mn}{(\gamma_J)_{\dot{m}}}^{p}\boldsymbol{k}_{I}
\end{eqnarray}
with the nilpotent generators given in terms of the $\boldsymbol{k}_{I}$ instead of $\boldsymbol{p}_{I}$. 

\subsection{Gaugings in $\mathbf{56}^{-1}$ and $\mathbf{8}^{+3}$ representations}
As an example for gaugings from an embedding tensor with more than one representation, we consider gaugings with both $\mathbf{56}^{-1}$ and $\mathbf{8}^{+3}$ representations. The embedding tensor takes the form
\begin{equation}\label{56+8theta}
\theta^{AM}\ = \ \left(\begin{array}{c|c|c}   & \theta_3^{m I} & \phantom{\theta_5^{m +}} \\\hline
								\theta_2^{\dot{m} -} &  &  \end{array}\right).
\end{equation}
Only $\theta_3^{m I}$ is constrained by the LC given in equation \eqref{56m_LC}. By a similar analysis as in the previous cases, we can solve this condition by parametrizing $\theta_3^{m I}$ as in equation \eqref{theta3PARA}. Denoting $\theta_2^{\dot{m} -}=v^{\dot{m}}$, we find that the QCs lead to the following four conditions, in accordance with the analysis of \cite{6D_11},
\begin{equation}
f_{\dot{m}\dot{n}\dot{p}}f^{\dot{m}\dot{n}\dot{p}}=0,\qquad f_{\dot{r}[\dot{m}\dot{n}}{f_{\dot{p}\dot{q}]}}^{\dot{r}}=0,\qquad f_{\dot{m}\dot{n}\dot{p}}v^{\dot{p}}=0,\qquad {f_{[\dot{m}\dot{n}\dot{p}}v_{\dot{q}]}|}_{SD}=0\label{QC_last}
\end{equation}
in which $|_{SD}$ means the self-dual part of a four-form. 
\\
\indent The first two conditions are independent of $v^{\dot{m}}$, and can be solved by the antisymmetric tensor $f_{\dot{m}\dot{n}\dot{p}}$ given in the case of gaugings from $\mathbf{56}^{-1}$ representation. We then consider the following three possibilities:
\begin{eqnarray}
 & & f_{\dot{m}\dot{n}\dot{p}}=\kappa(\varepsilon_{ijk},\varepsilon_{rst})\quad \text{for } i,j,...=\dot{1},\dot{2},\dot{3}\text{ and }r,s,...=\dot{5},\dot{6},\dot{7},\label{CSO401f}\\
 & & f_{\dot{m}\dot{n}\dot{p}}=\kappa(\varepsilon_{ijr},\varepsilon_{irs})\quad \text{for } i,j,...=\dot{1},\dot{2},\dot{3}\text{ and }r,s,...=\dot{5},\dot{6},\dot{7},\label{CSO311f}\\
 & & f_{\dot{m}\dot{n}\dot{p}}=\kappa(\varepsilon_{\bar{i}\bar{j}\bar{k}},\varepsilon_{\bar{r}\bar{s}\bar{t}})\quad \text{for }\bar{i},\bar{j},...=\dot{1},\dot{6},\dot{7}\text{ and }\bar{r},\bar{s},...=\dot{2},\dot{3},\dot{5}\label{CSO221f}
\end{eqnarray}
for a real constant $\kappa$. The last two conditions in \eqref{QC_last} can be solved by taking $v^{\dot{m}}$ with the only non-vanishing components given by 
\begin{equation}\label{vsoln}
v^{\dot{4}}=v^{\dot{8}}=\lambda
\end{equation}
for a real constant $\lambda$. With all these, the resulting gauge generators are given by
\begin{eqnarray}
X_m&=&(\gamma^I)_{m\dot{n}}v^{\dot{n}}\boldsymbol{k}_{I}+\frac{1}{24\sqrt{2}}f_{\dot{m}\dot{n}\dot{p}}(\gamma^{IJ})^{\dot{m}\dot{n}}{(\gamma_J)_m}^{\dot{p}}\boldsymbol{p}_{I},\\
X_{\dot{m}}&=&\frac{1}{48}f_{\dot{n}\dot{p}\dot{q}}(\gamma^{IK})^{\dot{n}\dot{p}}(\gamma_K)^{q\dot{q}}(\gamma^J)_{q\dot{m}}\boldsymbol{\tau}_{IJ}\, .
\end{eqnarray}
\indent It turns out that by a suitable redefinition of $X_{\dot{m}}$ generators, these generators can be shown to satisfy $SO(4-p,p)$ algebra
\begin{equation}\label{SOpqAlgebra}
\left[M_{\tilde{\mu}\tilde{\nu}},M_{\tilde{\rho}\tilde{\sigma}}\right]=2\kappa\left(\eta_{\tilde{\mu}[\tilde{\rho}}M_{\tilde{\sigma}]\tilde{\nu}}-\eta_{\tilde{\nu}[\tilde{\rho}}M_{\tilde{\sigma}]\tilde{\mu}}\right)
\end{equation}
with $\eta_{\tilde{\mu}\tilde{\nu}}=\text{diag}(1,1,1,1)$ for $f_{\dot{m}\dot{n}\dot{p}}$ in \eqref{CSO401f}, $\eta_{\tilde{\mu}\tilde{\nu}}=\text{diag}(1,1,1,-1)$ for $f_{\dot{m}\dot{n}\dot{p}}$ in \eqref{CSO311f}, and $\eta_{\tilde{\mu}\tilde{\nu}}=\text{diag}(1,1,-1,-1)$ for $f_{\dot{m}\dot{n}\dot{p}}$ in \eqref{CSO221f}. In addition, there are four constraints among $X_m$ generators, given by \eqref{irrCSO401TranCon}, implying that only four generators are linearly independent. Choosing these four generators as in section \ref{Sec56min}, we find the following commutation relations
\begin{eqnarray}
\left[K_{\tilde{\mu}},M_{\tilde{\nu}\tilde{\rho}}\right]&\hspace{-0.1cm}=&\hspace{-0.1cm}2\kappa\eta_{\tilde{\mu}[\tilde{\nu}}\delta_{\tilde{\rho}]}^{\tilde{\sigma}}K_{\tilde{\sigma}},\label{56KMcom}\\
\left[K_{\tilde{\mu}},K_{\tilde{\nu}}\right]&\hspace{-0.1cm}=&\hspace{-0.1cm}-\lambda\sqrt{2}M_{\tilde{\mu}\tilde{\nu}}\, .
\end{eqnarray}
It should be noted that in this case with non-vanishing $\lambda$, generators $K_{\tilde{\mu}}$ do not commute with each other but close onto the $SO(4-p,p)$ part. These generators enlarge $SO(4-p,p)$ to a larger gauge group. In particular, by setting $\lambda=\pm\frac{\kappa}{4\sqrt{2}}$ and defining the generators 
\begin{equation}\label{ExtendM}
M_{0\tilde{\mu}}=-M_{\tilde{\mu}0}=2K_{\tilde{\mu}},
\end{equation}
we obtain the algebra, with $\ul{\mu}=(0,\tilde{\mu})=0,1,2,3,4$,
\begin{equation}\label{simSO(PQ)algebra}
\left[M_{\ul{\mu}\ul{\nu}},M_{\ul{\rho}\ul{\sigma}}\right]=2\kappa\left(\eta_{\ul{\mu}[\ul{\rho}}M_{\ul{\sigma}]\ul{\nu}}-\eta_{\ul{\nu}[\ul{\rho}}M_{\ul{\sigma}]\ul{\mu}}\right)
\end{equation}
for $\eta_{\ul{\mu}\ul{\nu}}=\text{diag}(\pm1,\eta_{\tilde{\mu}\tilde{\nu}})$. The two sign choices correspond to $SO(5-p,p)$ or $SO(4-p,p+1)$ gauge groups. In particular, the $SO(4)$ group with $\eta_{\tilde{\mu}\tilde{\nu}}=\delta_{\tilde{\mu}\tilde{\nu}}$ is enlarged to $SO(5)$ or $SO(4,1)_{\textrm{I}}$. Similarly, the $SO(3,1)$ group with $\eta_{\tilde{\mu}\tilde{\nu}}=\text{diag}(1,1,1,-1)$ is enlarged to $SO(4,1)_{\textrm{II}}$ or $SO(3,2)_{\textrm{I}}$ while the $SO(2,2)$ group, with $\eta_{\tilde{\mu}\tilde{\nu}}=\text{diag}(1,1,-1,-1)$, becomes $SO(3,2)_{\textrm{II}}$. We have used subscripts I and II to distinguish the gauge groups arising from different $SO(4-p,p)$ groups obtained from the embedding tensor in $\mathbf{56}^{-1}$ representation. We also note that the analysis for gaugings from $\mathbf{56}^{+1}$ and $\mathbf{8}^{-3}$ representations can be carried out in the same way leading to the same gauge groups with the role of $X_m$ and $X_{\dot{m}}$ interchanged.
\\
\indent We end this section by some comments on the $SO(5-p,p)$ and $CSO(4-p,p,1)$ gauge groups identified in this section. The same gauge groups also arise in the classification of gauge groups under $GL(5)\subset SO(5,5)$ that has been extensively studied in \cite{our6DDW1}. In that case, both the $SO(5-p,p)$ and $CSO(4-p,p,1)$ gauge groups are embedded entirely in $GL(5)$ and are described by purely magnetic gaugings in which only components of the embedding tensor that couple the magnetic two-form fields ${B_{\mu\nu}}^{\overline{m}}$ are non-vanishing. Unlike the electric two-form fields $B_{\mu\nu\overline{m}}$, these fields are also accompanied by the three-form fields. On the other hand, the $CSO(4-p,p,1)$ and $SO(5-p,p)$ gauge groups are embedded respectively in $SO(4,4)$ and $SO(4,4)\ltimes \mathbb{R}^8$ with the $\mathbb{R}^8$ factor generated by $\boldsymbol{p}_I$ or $\boldsymbol{k}_I$ generators. As can be seen from the structure of the deformed $p$-form hierarchy given in \cite{6D_Max_Gauging}, the embedding tensor components $\mathbf{56}^{\pm 1}$ couple both electric and magnetic two-form fields $B_{\mu\nu\overline{m}}$ and ${B_{\mu\nu}}^{\overline{m}}$. Accordingly, the resulting gauged supergravities are not equivalent due to the different field contents among the tensor fields. In particular, gaugings obtained in \cite{our6DDW1} are known to arise from an $S^1$ reduction of $CSO(p,q,5-p-q)$ gauged supergravities in seven dimensions. However, higher dimensional origins of the gauge groups considered here are not clear at this stage. 
\section{Supersymmetric domain wall solutions}\label{DW_sol}
In this section, we find supersymmetric domain walls which are half-supersymmetric vacuum solutions of the maximal gauged supergravities considered in the previous section. We take the space-time metric to be the standard domain wall ansatz
\begin{equation}\label{DWmetric}
ds_6^2=e^{2A(r)}\eta_{\bar{\mu} \bar{\nu}}dx^{\bar{\mu}} dx^{\bar{\nu}}+dr^2
\end{equation}
where $\bar{\mu},\bar{\nu},\ldots $ are space-time indices of five-dimensional Minkowski space, and $A(r)$ is a warp factor depending only on the radial coordinate $r$. 
\\
\indent Following \cite{our6DDW1} and  \cite{our6DDW2}, the coset representative of $SO(5,5)/SO(5)\times SO(5)$, parametrized by 25 scalar fields, can be obtained by the following non-compact generators of $SO(5,5)$ in diagonal basis
\begin{equation}
\hat{\boldsymbol{t}}_{a\dot{b}}\,=\,{\mathbb{M}_{a}}^M{\mathbb{M}_{\dot{b}}}^N\boldsymbol{t}_{MN}
\end{equation}
where ${\mathbb{M}_{\underline{A}}}^M=({\mathbb{M}_{a}}^M,{\mathbb{M}_{\dot{a}}}^M)$ is the inverse of the transformation matrix $\mathbb{M}$ given in \eqref{offDiagTrans}. These non-compact generators are symmetric such that $(\hat{\boldsymbol{t}}_{a\dot{b}})^T=\hat{\boldsymbol{t}}_{a\dot{b}}$. Recall that an $SO(5,5)$ vector index in non-diagonal basis is decomposed as $M=(-,I,+)$ under $SO(4,4)$, we further decompose vector indices of both $SO(5)$ factors as $a=(0,i)$ and $\dot{a}=(\dot{i},\#)$, with $i=1,2,3,4$ and $\dot{i}=\dot{1},\dot{2},\dot{3},\dot{4}$. This leads to the following decomposition of the non-compact generators
\begin{equation}
\hat{\boldsymbol{t}}_{a\dot{b}}\,=\,(\hat{\boldsymbol{t}}_{0\dot{i}},\,\hat{\boldsymbol{t}}_{0\#},\,\hat{\boldsymbol{t}}_{i\dot{j}},\,\hat{\boldsymbol{t}}_{i\#}).
\end{equation}
\indent Under the $SO(4,4)$ branching rule for $SO(5,5)$ adjoint representation given in \eqref{SO55GenDec}, these non-compact generators read
\begin{eqnarray}
\hat{\boldsymbol{t}}_{0\dot{i}}&=&\frac{1}{\sqrt{2}}{\mathbb{M}_{\dot{i}}}^I(\boldsymbol{p}_{I}+\boldsymbol{k}_{I}),\qquad
\hat{\boldsymbol{t}}_{0\#}=-\boldsymbol{d},\nonumber\\
\hat{\boldsymbol{t}}_{i\#}&=&\frac{1}{\sqrt{2}}{\mathbb{M}_{i}}^I(\boldsymbol{p}_{I}-\boldsymbol{k}_{I}),\qquad
\hat{\boldsymbol{t}}_{i\dot{j}}={\mathbb{M}_{i}}^I{\mathbb{M}_{\dot{j}}}^J\boldsymbol{\tau}_{IJ}.
\end{eqnarray}
Under the compact $SO(5)\times SO(5)\subset SO(5,5)$, the $25$ scalars transform as $(\mathbf{5},\mathbf{5})$. The split of indices $a$ and $\dot{a}$ given above implies the branching $\mathbf{5}\rightarrow \mathbf{1}+\mathbf{4}$ of $SO(5)\rightarrow SO(4)$. Therefore, under $SO(4)\times SO(4)\subset SO(5)\times SO(5)$, the scalars tranform as
\begin{equation}\label{15repscalarDEC}
\underbrace{(\mathbf{5},\mathbf{5})}_{\hat{\boldsymbol{t}}_{a\dot{b}}}\ \rightarrow\ \underbrace{(\mathbf{1},\mathbf{1})}_{\boldsymbol{d}}\oplus\underbrace{(\mathbf{1},\mathbf{4})}_{\hat{\boldsymbol{t}}_{0\dot{i}}}\oplus\underbrace{(\mathbf{4},\mathbf{1})}_{\hat{\boldsymbol{t}}_{i\#}}\,\oplus\underbrace{(\mathbf{4},\mathbf{4})}_{{\boldsymbol{t}}_{i\dot{j}}}\, .
\end{equation}
\indent It is convenient to denote all $25$ scalar fields collectively as 
\begin{equation}
\Phi^{\mathbb{I}}=\{\varphi,\zeta_1,\ldots,\zeta_{4},\xi_{\dot{1}},\ldots,\xi_{\dot{4}},\phi_1,\ldots,\phi_{16}\}
\end{equation} 
with $\mathbb{I}=1,\ldots,25$. The scalar $\varphi$ is the dilaton corresponding to the $\mathbb{R}^+\sim SO(1,1)$ generator $\boldsymbol{d}$. The two sets of four scalars $\{\zeta_1,...,\zeta_{4}\}$ and $\{\xi_{\dot{1}},...,\xi_{\dot{4}}\}$ respectively correspond to the generators $\hat{\boldsymbol{t}}_{0\dot{i}}$ and $\hat{\boldsymbol{t}}_{i\#}$. The remaining sixteen scalar fields $\{\phi_1,...,\phi_{16}\}$ parametrize the submanifold $SO(4,4)/SO(4)\times SO(4)$ of the $SO(5,5)/SO(5)\times SO(5)$ coset. 
\\
\indent With this form of the scalar fields, we can rewrite the kinetic terms of the scalar fields in \eqref{bosonic_L} and obtain the following form of the bosonic Lagrangian
\begin{equation}
e^{-1}\mathcal{L}=\frac{1}{4}R-G_{{\mathbb{I}}{\mathbb{J}}}\partial_\mu\Phi^{\mathbb{I}}\partial^\mu\Phi^{\mathbb{J}}-\mathbf{V}\label{Ex_bosonic_L}
\end{equation}
with $G_{{\mathbb{I}}{\mathbb{J}}}=\frac{1}{16}P^{a\dot{a}}_{\mathbb{I}}P_{a\dot{a}\mathbb{J}}$ being a symmetric scalar metric. The vielbein $P^{a\dot{a}}_{\mathbb{I}}$ on the scalar manifold is related to $P^{a\dot{a}}_\mu$ via $P^{a\dot{a}}_{\mu}=P^{a\dot{a}}_{\mathbb{I}}\pd_\mu \Phi^{\mathbb{I}}$.
\\
\indent We will find supersymmetric domain wall solutions from first-order Bogomol'nyi-Prasad-Sommerfield (BPS) equations derived from the supersymmetry transformations of fermionic fields. The procedure is essentially the same as that given in \cite{our6DDW1} and  \cite{our6DDW2}, so we will mainly state the final results. The variations of the gravitini in \eqref{1stSUSY} and \eqref{2ndSUSY}, $\delta \psi_{+\bar{\mu}\alpha}$ and $\delta \psi_{-\bar{\mu}\dot{\alpha}}$ respectively gives
\begin{eqnarray} 
A'\hat{\gamma}_{r}\epsilon_{+\alpha}+\frac{1}{2}\Omega_{\alpha\beta}T^{\beta\dot{\alpha}}\epsilon_{-\dot{\alpha}}=0,\label{eq1_GBPS}\\
A'\hat{\gamma}_{r}\epsilon_{-\dot{\alpha}}-\frac{1}{2}\Omega_{\dot{\alpha}\dot{\beta}}T^{\alpha\dot{\beta}}\epsilon_{+\alpha}=0\, .\label{eq2_GBPS}
\end{eqnarray}
Throughout the paper, we use the notation $'$ to denote an $r$-derivative. Multiply equation \eqref{eq1_GBPS} by $A'\hat{\gamma}_{r}$ and use equation \eqref{eq2_GBPS} or vice-versa, we find the following consistency conditions
\begin{eqnarray}
{A'}^2{\delta_\alpha}^\beta &=&-\frac{1}{4}\Omega_{\alpha\gamma}T^{\gamma\dot{\alpha}}\Omega_{\dot{\alpha}\dot{\beta}}T^{\beta\dot{\beta}}=\mc{W}^2{\delta_\alpha}^\beta,\label{WarpBPSsq1}\\
{A'}^2{\delta_{\dot{\alpha}}}^{\dot{\beta}} &=&-\frac{1}{4}\Omega_{\dot{\alpha}\dot{\gamma}}T^{\alpha\dot{\gamma}}\Omega_{\alpha\beta}T^{\beta\dot{\beta}}
=\mc{W}^2{\delta_{\dot{\alpha}}}^{\dot{\beta}}\label{WarpBPSsq2}
\end{eqnarray}
in which we have introduced the ``superpotential'' $\mc{W}$. We then obtain the BPS equations for the warp factor
\begin{equation}
A'=\pm\mathcal{W}\, .\label{Ap_eq}
\end{equation}
With this result, equations \eqref{eq1_GBPS} and \eqref{eq2_GBPS} lead to the following (not independent) projectors on the Killing spinors
\begin{equation}\label{DW_Proj}
\hat{\gamma}_r\epsilon_{+\alpha}=-\frac{1}{2}\Omega_{\alpha\beta}\frac{T^{\beta\dot{\beta}}}{A'}\epsilon_{-\dot{\beta}},\qquad
\hat{\gamma}_r\epsilon_{-\dot{\alpha}}=\frac{1}{2}\Omega_{\dot{\alpha}\dot{\beta}}\frac{T^{\alpha\dot{\beta}}}{A'}\epsilon_{+\alpha}\, .
\end{equation}
Using these projectors in the variations $\delta\chi_{+a\dot{\alpha}}$ and $\delta\chi_{-\dot{a}\alpha}$ together with some identities involving the T-tensors, in particular \eqref{T_iden3}, we may rewrite the BPS equations for scalar fields in the form
\begin{equation}\label{BPSGenEq}
{\Phi^{\mathbb{I}}}'=\mp 2G^{\mathbb{I}\mathbb{J}}\frac{\partial\mathcal{W}}{\partial\Phi^{\mathbb{J}}}
\end{equation}
in which $G^{\mathbb{I}\mathbb{J}}$ is the inverse of the scalar metric $G_{\mathbb{I}\mathbb{J}}$. The remaining variations $\delta \psi_{+r\alpha}$ and $\delta\psi_{-r\dot{\alpha}}$ determine the $r$ dependence of the Killing spinors.
\\
\indent In addition, we also note that the scalar potential can be written in terms of $\mc{W}$ as
\begin{equation}\label{PopularSP}
\mathbf{V}=2G^{\mathbb{I}\mathbb{J}}\frac{\partial\mathcal{W}}{\partial\Phi^{\mathbb{I}}}\frac{\partial\mathcal{W}}{\partial\Phi^{\mathbb{J}}}-5\mathcal{W}^2\, .
\end{equation}
It is also straightforward to show that the BPS equations of the form \eqref{Ap_eq} and \eqref{BPSGenEq} satisfy the second-order field equations derived from the bosonic Lagrangian \eqref{Ex_bosonic_L} with the scalar potential given by \eqref{PopularSP}, see \cite{SPotWay1, SPotWay2, SPotWay3, SPotWay4, SPotWay5, SPotWay6} for more detail. Finally, since we have imposed only one independent projector on the Killing spinors, all solutions found in this work are half-supersymmetric. 
\subsection{$SO(4)$ symmetric domain walls}\label{56SO(4)soln}
To deal with the $25$-dimensional scalar manifold, we will follow the approach introduced in \cite{New_Extrema} by considering domain wall solutions that are invariant under a particular subgroup of the gauge groups. These solutions involve only a subset of all $25$ scalars. We also restrict ourselves to only gauge groups obtained from the embedding tensor in $\mathbf{56}^{-1}$ and $\mathbf{8}^{+3}$ representations.
\\
\indent We first consider supersymmetric domain walls with an unbroken symmetry $SO(4)$. Only $SO(5)$, $SO(4,1)_{\textrm{I}}$, and $CSO(4,0,1)$ gauge groups contain a common $SO(4)$ subgroup which in turn lies within $SO(4,4)$. We collectively describe them in a single framework by using $f_{\dot{m}\dot{n}\dot{p}}$ given in \eqref{CSO401f} together with $v^{\dot{4}}=v^{\dot{8}}=-\frac{\sigma\kappa}{4\sqrt{2}}$ with $\sigma=1,-1,$ and $0$, corresponding to $SO(5)$, $SO(4,1)_{\textrm{I}}$, and $CSO(4,0,1)$ gauge groups, respectively. The residual $SO(4)$ symmetry is embedded diagonally in $SO(4,4)$ via the maximal compact subgroup $SO(4)\times SO(4)$. From the decomposition of the scalars given in \eqref{15repscalarDEC}, we find two $SO(4)$ singlets according to the following decomposition
\begin{eqnarray}
(\mathbf{5},\mathbf{5})\rightarrow & &(\mathbf{1}\otimes \mathbf{1})\oplus(\mathbf{1}\otimes \mathbf{4})\oplus(\mathbf{4}\otimes\mathbf{1})\oplus(\mathbf{4}\otimes \mathbf{4})\nonumber \\
&\sim & \mathbf{1}\oplus\mathbf{4}\oplus\mathbf{4}\oplus\mathbf{1}\oplus\mathbf{6}\oplus\mathbf{9}\, .
\end{eqnarray}
The first singlet corresponds to the $\mathbb{R}^+$ generator $\boldsymbol{d}$ while the second one, arising from $\mathbf{4}\otimes \mathbf{4}$, is given by the non-compact generator
\begin{equation}\label{SO(4)singnoncom}
\mathcal{Y}=\hat{\boldsymbol{t}}_{1\dot{1}}-\hat{\boldsymbol{t}}_{2\dot{2}}-\hat{\boldsymbol{t}}_{3\dot{3}}-\hat{\boldsymbol{t}}_{4\dot{4}}
\end{equation}
from the $SO(4,4)/SO(4)\times SO(4)$ coset.
\\
\indent Using the coset representative of the form
\begin{equation}\label{SO4coset}
V=e^{\varphi\boldsymbol{d}+\phi\mathcal{Y}},
\end{equation}
we find the superpotential and the scalar potential of the form
\begin{eqnarray}
\mathcal{W}&\hspace{-0.1cm}=&\hspace{-0.1cm}\frac{g \kappa}{16 \sqrt{2}}  e^{-3 \varphi-4 \phi} \left(4 e^{4 (\varphi+\phi)}+\sigma\right),\label{SO(4)SPot}\\
\mathbf{V}&\hspace{-0.1cm}=&\hspace{-0.1cm}-\frac{g^2 \kappa ^2}{64}e^{-6 \varphi-8 \phi} \left(8 e^{8 (\varphi+\phi)}+8 \sigma  e^{4 (\varphi+\phi)}-\sigma^2\right).\qquad\label{SO(4)Pot}
\end{eqnarray}
It can be checked that the scalar potential can be written in terms of the superpotential according to \eqref{PopularSP} using the scalar matrix $G^{\mathbb{I}\mathbb{J}}=\text{diag}(\frac{1}{2}, \frac{1}{8})$ for $\Phi^{\mathbb{I}}=\{\varphi,\phi\}$ and $\mathbb{I}=1,2$. The general analysis given above leads to the BPS equation for the warp factor
\begin{equation}\label{SO(4)Aprim}
A'=\pm\frac{g \kappa}{16 \sqrt{2}}  e^{-3 \varphi-4 \phi} \left(4 e^{4 (\varphi+\phi)}+\sigma\right)
\end{equation}
together with the BPS equations for the scalar fields
\begin{eqnarray}\label{SO(4)phiprim}
& &\varphi'=\mp\frac{g \kappa}{16 \sqrt{2}}  e^{-3 \varphi-4 \phi} \left(4 e^{4 (\varphi+\phi)}-3\sigma\right)\nonumber \\
\text{ and }\qquad & &\phi'=\pm\frac{\sigma g \kappa}{16 \sqrt{2}}  e^{-3 \varphi-4 \phi}\, .
\end{eqnarray}
\indent For $\sigma=1,-1$ corresponding to $SO(5)$ and $SO(4,1)_{\textrm{I}}$ gauge groups, the solutions for the warp factor $A$ and dilaton $\varphi$ can be given in terms of $\phi$ as
\begin{eqnarray}
A&\hspace{-0.1cm}=&\hspace{-0.1cm}C_1+3\phi-\frac{1}{4} \ln \left[e^{-4 (4 \phi+C_2)}+\sigma\right],\label{SO(4)Asoln}\\
\varphi&\hspace{-0.1cm}=&\hspace{-0.1cm}-\phi-\frac{1}{4} \ln \left[e^{-4 (4 \phi+C_2)}+\sigma\right]\label{SO(4)phisoln}
\end{eqnarray}
in which $C_1$ and $C_2$ are integration constants. To obtain the solution for $\phi$, we change the radial coordinate $r$ to $\rho$ defined by $\frac{d\rho}{dr}=e^{-3 \varphi-4 \phi}$. The solution of $\phi$ is then readily found to be
\begin{equation}
\phi=\pm\frac{g\sigma\kappa\rho}{16\sqrt{2}}+C_3\label{SO4_phi_sol_final}
\end{equation}
in which $\pm$ directly corresponds to the upper/lower signs in the BPS equations. Thus, the two sign choices in the BPS equations can be absorbed by flipping the sign of the radial coordinate. We will neglect these sign choices by choosing the upper sign of the BPS equations from now on. Moreover, the integration constants $C_1$ and $C_3$ can also be removed by rescaling the coordinates $x^{\bar\mu}$ and shifting the radial coordinate $\rho$.
\\
\indent For $\sigma=0$ corresponding to $CSO(4,0,1)$ gauge group, the superpotential and scalar potential are independent of $\phi$  
\begin{equation}\label{CSO(401)Pot}
\mathcal{W}=\frac{g \kappa}{4 \sqrt{2}}  e^\varphi\qquad\text{ and }\qquad \mathbf{V}=-\frac{g^2 \kappa ^2}{8}e^{2\varphi},
\end{equation}
and the BPS equations reduce to
\begin{equation}
A'=-\varphi'=\frac{g \kappa}{4 \sqrt{2}} e^\varphi\qquad\text{ and }\qquad\phi'=0.
\end{equation}
All of these equations can be readily solved to obtain the solution
\begin{equation}\label{YSO(5)DW}
A=-\varphi=\ln\left(\frac{g\kappa r}{4\sqrt{2}}-C\right)\qquad \text{ and }\qquad \phi=0\, .
\end{equation}
In this case, we can consistently truncate out the $SO(4)$ invariant scalar $\phi$ since the scalar potential is independent of $\phi$.
\\
\indent For $SO(4,1)_{\textrm{II}}$ gauge group, the $SO(4)$ compact subgroup is not embedded in $SO(4,4)$ since it involves $M_{0\tilde{i}}$ generators obtained from the gauge generators $X_m$. However, a similar analysis can be carried out by using $f_{\dot{m}\dot{n}\dot{p}}$ given in \eqref{CSO311f} together with $v^{\dot{4}}=v^{\dot{8}}=\frac{\kappa}{4\sqrt{2}}$. In this case, there are again two $SO(4)$ singlet scalars corresponding to the non-compact generators
\begin{equation}
\overline{\mathbb{Y}}_1=\hat{\boldsymbol{t}}_{1\dot{1}}\qquad\text{ and }\qquad \overline{\mathbb{Y}}_2=\boldsymbol{d}-\hat{\boldsymbol{t}}_{2\dot{2}}-\hat{\boldsymbol{t}}_{3\dot{3}}
-\hat{\boldsymbol{t}}_{4\dot{4}}\, .
\end{equation}
Using the coset representative 
\begin{equation}
V=e^{\overline{\phi}_1\overline{\mathbb{Y}}_1+\overline{\phi}_2\overline{\mathbb{Y}}_1},
\end{equation}
we find the same form of the domain wall solution as given in \eqref{SO(4)Asoln}-\eqref{SO4_phi_sol_final} with $\sigma=-3$, $\varphi=-\overline{\phi}_1$, and $\phi=-\overline{\phi}_2$.

\subsection{$SO(3)$ symmetric domain walls}
We next consider supersymmetric domain walls preserving a smaller residual symmetry $SO(3)$ generated by $M_{ij}$ for $i,j,...=1,2,3$ from the $SO(4-p,p)$ generators $M_{\tilde{\mu}\tilde{\nu}}$. There are many gauge groups containing the $SO(3)\subset SO(4,4)$. These are given by $SO(5)$, $SO(4,1)_{\textrm{I}}$, and $CSO(4,0,1)$ with $f_{\dot{m}\dot{n}\dot{p}}$ given in \eqref{CSO401f} and $v^{\dot{4}}=v^{\dot{8}}=-\frac{\sigma\kappa}{4\sqrt{2}}$ together with $SO(4,1)_{\textrm{II}}$, $SO(3,2)_{\textrm{I}}$, and $CSO(3,1,1)$ with $f_{\dot{m}\dot{n}\dot{p}}$ given in \eqref{CSO311f} and $v^{\dot{4}}=v^{\dot{8}}=\frac{\sigma\kappa}{4\sqrt{2}}$.
\\
\indent By further decomposing the residual symmetry of the previous section according to $SO(4)\rightarrow SO(3)$ with the branching $\mathbf{4}\rightarrow \mathbf{1}+\mathbf{3}$, we find the following transformation of all $25$ scalars
\begin{eqnarray}
(\mathbf{5},\mathbf{5})\rightarrow& &(\mathbf{1}\oplus\mathbf{1}\oplus\mathbf{3})\otimes (\mathbf{1}\oplus\mathbf{1}\oplus\mathbf{3})\nonumber \\
&\sim&4(\mathbf{1})\oplus4(\mathbf{3})\oplus(\mathbf{1}\oplus\mathbf{3}\oplus\mathbf{5})
\end{eqnarray}
with the last three representations arising from the product $\mathbf{3}\otimes \mathbf{3}$. There are then five scalars invariant under the residual symmetry $SO(3)$. One of them is again the dilaton corresponding to the $\mathbb{R}^+$ generator. The other four singlets are associated with the following non-compact generators 
\begin{equation}
\mc{Y}_1=\hat{\boldsymbol{t}}_{1\dot{1}},\qquad \mc{Y}_2=\hat{\boldsymbol{t}}_{2\dot{2}}+\hat{\boldsymbol{t}}_{3\dot{3}}+\hat{\boldsymbol{t}}_{4\dot{4}},\qquad \mc{Y}_3=\hat{\boldsymbol{t}}_{0\dot{1}},\qquad \mc{Y}_4=\hat{\boldsymbol{t}}_{1\#}\, .
\end{equation}
We can use the coset representative of the form
\begin{equation}
V=e^{\varphi\boldsymbol{d}+\phi_1\mc{Y}_1+\phi_2\mc{Y}_2+\zeta\mc{Y}_3+\xi\mc{Y}_4}\, .
\end{equation}
\indent It turns out that consistency between the resulting BPS equations and the second-order field equations requires $\xi=\zeta$. With this condition, the superpotential and scalar potential become
\begin{eqnarray}
\mathcal{W}&\hspace{-0.2cm}=&\hspace{-0.2cm}\frac{g \kappa }{16 \sqrt{2}}e^{-(3\varphi+\phi_1+\phi_2)}\left[3e^{4\varphi}+e^{4 \phi_2}\cosh{4\zeta}\left(ve^{4 (\varphi+\phi_1)}+\sigma \right)\right],\\
\mathbf{V}&\hspace{-0.2cm}=&\hspace{-0.2cm}\frac{1}{64}e^{-2(3\varphi+\phi_1+\phi_2)}g^2 \kappa ^2\left[\left[e^{4\phi_2} \cosh{4\zeta}\left(ve^{4 (\varphi+\phi_1)}+\sigma\right)-3 e^{4\varphi}\right]^2\right. \nonumber \\
& &\left.-12 e^{8 \varphi}-4v \sigma e^{4 (\varphi+\phi_1+2\phi_2)}\right].\qquad\quad
\end{eqnarray}
We have used the parameter $v=2u-1$ with $u=\pm1$ together with $\sigma=\pm1,0$ to identify the gauge groups; $SO(5)$ ($u=\sigma=1$), $SO(4,1)_{\textrm{I}}$ ($u=-\sigma=1$), $CSO(4,0,1)$ ($u=1, \sigma=0$), $SO(4,1)_{\textrm{II}}$ ($u=-\sigma=-1$), $SO(3,2)_{\textrm{I}}$ ($u=\sigma=-1$), and $CSO(3,1,1)$ ($u=-1,\sigma=0$). 
\\
\indent Using
\begin{equation}
G^{\mathbb{I}\mathbb{J}}=\frac{1}{8}\begin{pmatrix} (3+\cosh{8\zeta})\text{sech}^2{4\zeta} & -2\tanh^2{4\zeta} &0 &0 \\
								-2\tanh^2{4\zeta} & (3+\cosh{8\zeta})\text{sech}^2{4\zeta} &0 &0\\
									0 & 0 & \frac{4}{3} & 0 \\ 0 & 0 & 0 & 2
 \end{pmatrix}
\end{equation}
for $\Phi^{\mathbb{I}}=\{\varphi,\phi_1,\phi_2,\zeta\}$ with $\mathbb{I}=1,2,3,4$, we can rewrite the scalar potential in terms of the superpotential as in \eqref{PopularSP}. 
\\
\indent With all these, we arrive at the BPS equations
\begin{eqnarray}
A'&\hspace{-0.2cm}=&\hspace{-0.2cm}\frac{g \kappa }{16 \sqrt{2}}e^{-(3\varphi+\phi_1+\phi_2)}\left[3e^{4\varphi}+e^{4 \phi_2}\cosh{4\zeta}\left(ve^{4 (\varphi+\phi_1)}+\sigma \right)\right],\\
\varphi'&\hspace{-0.2cm}=&\hspace{-0.2cm}-\frac{1}{32 \sqrt{2}} e^{-3\varphi-\phi_1-\phi_2}g\kappa\left[6e^{4\varphi}\right. \nonumber \\
& &\left.-e^{4 \phi_2}\text{sech}{4\zeta}\left[v e^{4 (\varphi+\phi_1)}(\cosh{8\zeta}-3)+\sigma(\cosh{8\zeta}+5) \right]\right],\qquad\ \\
\phi'_1&\hspace{-0.2cm}=&\hspace{-0.2cm}\frac{1}{32 \sqrt{2}} e^{-3\varphi-\phi_1-\phi_2}g\kappa\left[6e^{4\varphi}\right. \nonumber \\
& &\left.-e^{4 \phi_2}\text{sech}{4\zeta}\left[v e^{4 (\varphi+\phi_1)}(\cosh{8\zeta}+5)+\sigma(\cosh{8\zeta}-3) \right]\right],\\
\phi'_2&\hspace{-0.2cm}=&\hspace{-0.2cm}\frac{g\kappa} {16 \sqrt{2}}e^{-(3\varphi+\phi_1+\phi_2)}\left[e^{4\varphi}-e^{4 \phi_2}\cosh{4\zeta}\left(ve^{4 (\varphi+\phi_1)}+\sigma \right) \right],\\
\zeta'&\hspace{-0.2cm}=&\hspace{-0.2cm}-\frac{g \kappa}{8 \sqrt{2}}e^{-3\varphi-\phi_1+3\phi_2}\sinh{4\zeta} \left(ve^{4 (\varphi+\phi_1)}+\sigma \right).
\end{eqnarray}
From these equations, the solutions for scalar fields $\phi_2$, $\phi_1$, $\varphi$, and the warp factor $A$ can be obtained as functions of $\zeta$. These are given by
\begin{eqnarray}
\phi_2&\hspace{-0.2cm}=&\hspace{-0.2cm}\frac{1}{16} \ln \left[\frac{\sinh{8\zeta}\sqrt{v \sigma+C_1^2-C_1^2 \text{sech}^2{4\zeta}}-C_2v^2\sigma^2\sinh ^2{4\zeta}}{2 v \sigma}\right],\\
\phi_1&\hspace{-0.2cm}=&\hspace{-0.2cm}\frac{1}{4} \ln \left[\frac{\sqrt{C_1^2+v\sigma-C_1^2\text{sech}^2{4\zeta}}-C_1 \tanh{4\zeta}}{v}\right]-\frac{1}{4}\ln{\sinh{4\zeta}}+3\phi_2,\qquad\\
\varphi&\hspace{-0.2cm}=&\hspace{-0.2cm}\frac{1}{4} \ln \left[\frac{\sqrt{C_1^2+v\sigma-C_1^2\text{sech}^2{4\zeta}}-C_1 \tanh{4\zeta}}{v}\right]+\frac{1}{4}\ln{\sinh{4\zeta}}-3\phi_2,\qquad\\
A&\hspace{-0.2cm}=&\hspace{-0.2cm}3\phi_2-\frac{1}{2}\ln{\sinh{4\zeta}}
\end{eqnarray}        
in which we have chosen the integration constants for $\varphi$ and $A$ to be zero for simplicity. To obtain the solution for $\zeta$, we change $r$ to a new radial coordinate $\rho$ defined by $\frac{d\rho}{dr}=e^{-4\zeta-\varphi-\phi_1+3\phi_2}(v e^{4 (\varphi+\phi_1)}+\sigma)$. The solution for $\zeta$ is then given by
\begin{equation}
e^{4\zeta}=\tanh\left(\frac{g\kappa\rho}{4\sqrt{2}}-C_3\right).
\end{equation}     
\indent It should be noted that for the case of $SO(3,2)_{\textrm{II}}$ gauge group, the $SO(3)$ subgroup lies outside $SO(4,4)$. A similar solution with $v=\sigma=-1$ can be found by using $f_{\dot{m}\dot{n}\dot{p}}$ given in \eqref{CSO221f} together with $v^{\dot{4}}=v^{\dot{8}}=\frac{\kappa}{4\sqrt{2}}$, and the coset representative
\begin{equation}
V=e^{-\varphi\hat{\boldsymbol{t}}_{1\dot{1}}-\phi_1\hat{\boldsymbol{t}}_{4\dot{4}}
+\phi_2(\boldsymbol{d}+\hat{\boldsymbol{t}}_{2\dot{2}}+\hat{\boldsymbol{t}}_{3\dot{3}})+\zeta(\hat{\boldsymbol{t}}_{1\dot{4}}-\hat{\boldsymbol{t}}_{4\dot{1}})}
\end{equation}
with the five $SO(3)$ singlets corresponding to the non-compact generators $\hat{\boldsymbol{t}}_{1\dot{1}}$, $\hat{\boldsymbol{t}}_{4\dot{4}}$, $\boldsymbol{d}+\hat{\boldsymbol{t}}_{2\dot{2}}+\hat{\boldsymbol{t}}_{3\dot{3}}$, $\hat{\boldsymbol{t}}_{1\dot{4}}$, and $\hat{\boldsymbol{t}}_{4\dot{1}}$. We also note the relation between scalar fields corresponding to $\hat{\boldsymbol{t}}_{1\dot{4}}$ and $\hat{\boldsymbol{t}}_{4\dot{1}}$ generators arising from consistency between the BPS equations and the field equations as in the above analysis.
\subsection{$SO(2)\times SO(2)$ symmetric domain walls}
We now move to domain wall solutions with $SO(2)\times SO(2)$ symmetry. Gauge groups containing an $SO(2)\times SO(2)$ subgroup, embedded entirely in $SO(4,4)$, are $SO(5)$, $SO(4,1)_{\textrm{I}}$, $CSO(4,0,1)$ with $f_{\dot{m}\dot{n}\dot{p}}$ given in \eqref{CSO401f} and $v^{\dot{4}}=v^{\dot{8}}=-\frac{\sigma\kappa}{4\sqrt{2}}$ together with $SO(3,2)_{\textrm{II}}$ and $CSO(2,2,1)$ with $f_{\dot{m}\dot{n}\dot{p}}$ given in \eqref{CSO221f} and $v^{\dot{4}}=v^{\dot{8}}=\frac{\sigma\kappa}{4\sqrt{2}}$. In order to incorporate the two sets of gauge groups within a single framework, we will choose the $SO(2)\times SO(2)$ generators to be $X_{\dot{1}}$ and $X_{\dot{5}}$.
\\
\indent The residual symmetry $SO(2)\times SO(2)$ is embedded in $SO(4)$ as $\mathbf{4}\rightarrow (\mathbf{1},\mathbf{2})+(\mathbf{2},\mathbf{1})$ with $\mathbf{2}$ denoting the fundamental or vector representation of $SO(2)$. As in the previous cases, decomposing the transformation of all $25$ scalar fields under $SO(2)\times SO(2)$ gives
\begin{eqnarray}
(\mathbf{5},\mathbf{5})\rightarrow& &[(\mathbf{1},\mathbf{1})\oplus(\mathbf{1},\mathbf{2})\oplus(\mathbf{2},\mathbf{1})]\otimes [(\mathbf{1},\mathbf{1})\oplus(\mathbf{1},\mathbf{2})\oplus(\mathbf{2},\mathbf{1})]\nonumber \\
&\sim&5(\mathbf{1},\mathbf{1})\oplus3(\mathbf{1},\mathbf{2})\oplus3(\mathbf{2},\mathbf{1})\oplus2(\mathbf{2},\mathbf{2}).\label{SO2_SO2_decom}
\end{eqnarray}
Accordingly, there are five singles corresponding to the $\mathbb{R}^+$ generator $\boldsymbol{d}$ and the following four non-compact generators
\begin{equation}\label{SO2xSO2noncom}
\mathds{Y}_1=\hat{\boldsymbol{t}}_{1\dot{1}}-\hat{\boldsymbol{t}}_{4\dot{4}},\qquad
\mathds{Y}_2=\hat{\boldsymbol{t}}_{2\dot{2}}+\hat{\boldsymbol{t}}_{3\dot{3}},\qquad
\mathds{Y}_3=\hat{\boldsymbol{t}}_{1\dot{4}}+\hat{\boldsymbol{t}}_{4\dot{1}},\qquad
\mathds{Y}_4=\hat{\boldsymbol{t}}_{2\dot{3}}-\hat{\boldsymbol{t}}_{3\dot{2}}\, .
\end{equation}
leading to the coset representative of the form
\begin{equation}
V=e^{\varphi\boldsymbol{d}+\phi_1\mathds{Y}_1+\phi_2\mathds{Y}_2+\phi_3\mathds{Y}_3+\phi_4\mathds{Y}_4}\, .
\end{equation}
\indent In this case, consistency between the BPS equations and field equations requires vanishing $\phi_3$ and $\phi_4$. With $\phi_3=\phi_4=0$, the superpotential is given by
\begin{equation}
\mathcal{W}=\frac{g \kappa}{16\sqrt{2}}e^{\varphi-2 (\phi_1+\phi_2)}  \left(2e^{4 (\phi_1+\phi_2)}+2u+u\sigma e^{4(\phi_2-\varphi)}\right),
\end{equation}
and the scalar potential takes the form
\begin{equation}
\mathbf{V}=-\frac{ug^2 \kappa ^2}{64}e^{-6\varphi-4\phi_1} \left[8 e^{8 \varphi+4 \phi_1}+4 \sigma e^{4 \varphi}( e^{4 (\phi_1+\phi_2)}+u)-u\sigma^2 e^{4 \phi_2}\right].
\end{equation}
The parameters $u=\pm1$ and $\sigma=\pm1,0$ correspond to different gauge groups; $SO(5)$ ($u=\sigma=1$), $SO(4,1)_{\textrm{I}}$ ($u=-\sigma=1$), $CSO(4,0,1)$ ($u=1, \sigma=0$), $SO(3,2)_{\textrm{II}}$ ($u=-\sigma=-1$ or $u=\sigma=-1$), and $CSO(2,2,1)$ ($u=-1,\sigma=0$).
\\
\indent The resulting BPS equations read
\begin{eqnarray}
A'&\hspace{-0.1cm}=&\hspace{-0.1cm}\frac{g \kappa}{16\sqrt{2}}e^{\varphi-2 (\phi_1+\phi_2)}  \left(2e^{4 (\phi_1+\phi_2)}+2u+u\sigma e^{4(\phi_2-\varphi)}\right),\\
\varphi'&\hspace{-0.1cm}=&\hspace{-0.1cm}-\frac{g \kappa}{16\sqrt{2}}e^{\varphi-2 (\phi_1+\phi_2)}  \left(2e^{4 (\phi_1+\phi_2)}+2u-3u\sigma e^{4(\phi_2-\varphi)}\right),\\
\phi'_1&\hspace{-0.1cm}=&\hspace{-0.1cm}-\frac{g \kappa}{16\sqrt{2}}e^{\varphi-2 (\phi_1+\phi_2)}  \left(2e^{4 (\phi_1+\phi_2)}-2u-u\sigma e^{4(\phi_2-\varphi)}\right),\\
\phi'_2&\hspace{-0.1cm}=&\hspace{-0.1cm}-\frac{g \kappa}{16\sqrt{2}}e^{\varphi-2 (\phi_1+\phi_2)}  \left(2e^{4 (\phi_1+\phi_2)}-2u+u\sigma e^{4(\phi_2-\varphi)}\right).
\end{eqnarray}
Defining a new radial coordinate $\rho$ by $\frac{d\rho}{dr}=e^{\varphi-2 (\phi_1+\phi_2)}$, we find a domain wall solution
\begin{eqnarray}
\phi_2&\hspace{-0.1cm}=&\hspace{-0.1cm}C_1+\frac{1}{16}\ln\left[e^{C_3}+\sigma  e^{\frac{ug \kappa  \rho}{\sqrt{2}}}\right]+\frac{1}{8}\ln \left[1-e^{u(C_2-\frac{g \kappa  \rho}{\sqrt{2}})}\right],\\
\phi_1&\hspace{-0.1cm}=&\hspace{-0.1cm}-\phi_2-\frac{1}{4}\ln \left[u(1-e^{u(C_2-\frac{g \kappa  \rho}{\sqrt{2}})})\right],\\
\varphi&\hspace{-0.1cm}=&\hspace{-0.1cm}\phi_2+\frac{1}{4}\ln \left[e^{C_3-\frac{ug \kappa  \rho}{\sqrt{2}}}+\sigma\right],\\
A&\hspace{-0.1cm}=&\hspace{-0.1cm}-5\phi_2-\frac{1}{2}\ln \left[u(1-e^{u(C_2-\frac{g \kappa  \rho}{\sqrt{2}})})\right]-\frac{1}{4}\ln \left[e^{C_3-\frac{ug \kappa  \rho}{\sqrt{2}}}+\sigma\right]
\end{eqnarray}
with the integration constants $C_1$, $C_2$, and $C_3$. For simplicity, we have chosen an integration constant for $A$ to be zero. As in the previous cases, by suitably redefining $u$ and $v$ parameters together with $SO(2)\times SO(2)$ singlet scalars, we can find a similar solution for $SO(3,2)_{\textrm{I}}$ gauge group in which the $SO(2)\times SO(2)$ residual symmetry lies outside the $SO(4,4)$. 
\subsection{$SO(2)_{\text{diag}}$ symmetric domain walls}
We finally consider supersymmetric domain walls preserving an $SO(2)_{\text{diag}}\subset SO(2)\times SO(2)$ symmetry. 
By taking the product among the various representations in \eqref{SO2_SO2_decom} to implement the $SO(2)_{\text{diag}}$ subgroup, we find
\begin{eqnarray}
(\mathbf{5},\mathbf{5})\rightarrow 9(\mathbf{1})\oplus8(\mathbf{2})
\end{eqnarray}
leading to nine singlet scalars. Among these singlets, five of them correspond to the $SO(2)\times SO(2)$ singlets given in the previous section. The additional four singlets correspond to the following non-compact generators
\begin{equation}\label{ADDSO2dnoncom}
\mathds{Y}_5=\hat{\boldsymbol{t}}_{1\dot{2}}-\hat{\boldsymbol{t}}_{4\dot{3}},\qquad
\mathds{Y}_6=\hat{\boldsymbol{t}}_{2\dot{1}}+\hat{\boldsymbol{t}}_{3\dot{4}},\qquad
\mathds{Y}_7=\hat{\boldsymbol{t}}_{1\dot{3}}+\hat{\boldsymbol{t}}_{4\dot{2}},\qquad
\mathds{Y}_8=\hat{\boldsymbol{t}}_{2\dot{4}}-\hat{\boldsymbol{t}}_{3\dot{1}}\, .
\end{equation}
The coset representative can be written as
\begin{equation}
V=e^{\varphi\boldsymbol{d}+\phi_1\mathds{Y}_1+\phi_2\mathds{Y}_2+\phi_3\mathds{Y}_3+\phi_4\mathds{Y}_4+\phi_5\mathds{Y}_5+\phi_6\mathds{Y}_6+\phi_7\mathds{Y}_7+\phi_8\mathds{Y}_8}\, .
\end{equation} 
It turns out that the resulting T-tensor, superpotential, and scalar potential are highly complicated. Accordingly, we will look for some subtruncations to simplify the analysis but still obtain interesting results. One possibility is to impose the conditions $\phi_3=\phi_4=0$ together with $\phi_8=\phi_7$ and $\phi_6=-\phi_5$. We have checked that these indeed lead to a consistent subtruncation.
\\
\indent The truncated coset representative is now given by
\begin{equation}
V=e^{\varphi\boldsymbol{d}+\phi_1\mathds{Y}_1+\phi_2\mathds{Y}_2+\tilde{\phi}_3(\mathds{Y}_5-\mathds{Y}_6)+\tilde{\phi}_4(\mathds{Y}_7+\mathds{Y}_8)}
\end{equation} 
giving rise to the superpotential and scalar potential
\begin{eqnarray}
\mathcal{W}&\hspace{-0.2cm}=&\hspace{-0.2cm}\frac{g \kappa}{16\sqrt{2}}e^{\varphi-2 (\phi_1+\phi_2)}\left[(2e^{4 (\phi_1+\phi_2)}+2u)\cosh{4\tilde{\phi}_3}\cosh{4\tilde{\phi}_4}+u\sigma e^{4(\phi_2-\varphi)}\right],\qquad \\
\mathbf{V}&\hspace{-0.2cm}=&\hspace{-0.2cm}-\frac{1}{64}e^{-6\varphi-4\phi_1}ug^2 \kappa ^2 \left[4 \sigma e^{4 \varphi}( e^{4 (\phi_1+\phi_2)}+u)\cosh{4\tilde{\phi}_3}\cosh{4\tilde{\phi}_4}\right. \nonumber \\
& &\left.+8 e^{8 \varphi4 \phi_1}-u\sigma^2 e^{4 \phi_2}\right].\quad\nonumber\\
\end{eqnarray}
The latter can be written in terms of the former according to \eqref{PopularSP} using
\begin{equation}
G^{\mathbb{I}\mathbb{J}}=\frac{1}{8}\begin{pmatrix} 4 & 0 & 0 & 0 & 0 \\ 0 & \text{sech}^24\tilde{\phi_3}\,\text{sech}^24\tilde{\phi_4}+1 & \text{sech}^24\tilde{\phi_3}\,\text{sech}^24\tilde{\phi_4}-1 & 0 & 0 \\ 0 & \text{sech}^24\tilde{\phi_3}\,\text{sech}^24\tilde{\phi_4}-1 & \text{sech}^24\tilde{\phi_3}\,\text{sech}^24\tilde{\phi_4}+1 & 0 & 0 \\ 0 & 0 & 0 & \text{sech}^24\tilde{\phi_4} & 0 \\ 0 & 0 & 0 & 0 & 1 \\\end{pmatrix}
\end{equation}
for $\Phi^{\mathbb{I}}=\{\varphi, \phi_1, \phi_2, \tilde{\phi}_3, \tilde{\phi}_4\}$ with $\mathbb{I},\mathbb{J},\ldots =1,2,3,4,5$. 
\\
\indent In this case, various gauge groups are identified by the parameters $u=\pm1$ and $\sigma=\pm1,0$ as $SO(5)$ ($u=\sigma=1$), $SO(4,1)_{\textrm{I}}$ ($u=-\sigma=1$), $CSO(4,0,1)$ ($u=1, \sigma=0$), $SO(3,2)_{\textrm{II}}$ ($u=-\sigma=-1$ or $u=\sigma=-1$), and $CSO(2,2,1)$ ($u=-1,\sigma=0$). With all these, we find the BPS equations
\begin{eqnarray}
A'&\hspace{-0.2cm}=&\hspace{-0.2cm}\frac{g \kappa}{16\sqrt{2}}e^{\varphi-2 (\phi_1+\phi_2)}\left[(2e^{4 (\phi_1+\phi_2)}+2u)\cosh{4\tilde{\phi}_3}\cosh{4\tilde{\phi}_4}+u\sigma e^{4(\phi_2-\varphi)}\right],\\
\varphi'&\hspace{-0.2cm}=&\hspace{-0.2cm}-\frac{g \kappa}{16\sqrt{2}}e^{\varphi-2 (\phi_1+\phi_2)}\left[(2e^{4 (\phi_1+\phi_2)}+2u)\cosh{4\tilde{\phi}_3}\cosh{4\tilde{\phi}_4}-3u\sigma e^{4(\phi_2-\varphi)}\right],\qquad\ \\
\phi'_1&\hspace{-0.2cm}=&\hspace{-0.2cm}-\frac{g \kappa}{16\sqrt{2}}e^{\varphi-2 (\phi_1+\phi_2)}\left[(2e^{4 (\phi_1+\phi_2)}-2u)\text{sech}{4\tilde{\phi}_3}\,\text{sech}{4\tilde{\phi}_4}-u\sigma e^{4(\phi_2-\varphi)}\right],\\
\phi'_2&\hspace{-0.2cm}=&\hspace{-0.2cm}-\frac{g \kappa}{16\sqrt{2}}e^{\varphi-2 (\phi_1+\phi_2)}\left[(2e^{4 (\phi_1+\phi_2)}-2u)\text{sech}{4\tilde{\phi}_3}\,\text{sech}{4\tilde{\phi}_4}+u\sigma e^{4(\phi_2-\varphi)}\right],\\
\tilde{\phi}'_3&\hspace{-0.2cm}=&\hspace{-0.2cm}-\frac{g \kappa}{16\sqrt{2}}e^{\varphi-2 (\phi_1+\phi_2)}(2e^{4 (\phi_1+\phi_2)}+2u)\sinh{4\tilde{\phi}_3}\,\text{sech}{4\tilde{\phi}_4},\\
\tilde{\phi}'_4&\hspace{-0.2cm}=&\hspace{-0.2cm}-\frac{g \kappa}{16\sqrt{2}}e^{\varphi-2 (\phi_1+\phi_2)}(2e^{4 (\phi_1+\phi_2)}+2u)\cosh{4\tilde{\phi}_3}\sinh{4\tilde{\phi}_4}.
\end{eqnarray}
Changing the radial coordinate to $\rho$ defined by $\frac{d\rho}{dr}=e^{\varphi-2 (\phi_1+\phi_2)}$, we can solve these equations to find a domain wall solution
\begin{eqnarray}
\tilde{\phi}_3&\hspace{-0.1cm}=&\hspace{-0.1cm}\frac{1}{4} \sinh ^{-1}\left[\frac{4C_5 \sqrt{u}}{\sqrt{C_0^6 g^2 \kappa ^2 \left(C_6+2u\sigma g^3 \kappa ^3 \rho ^4\right)^2-16 C_5^2 \left(C_1^2+u\right)}}\right],\\
\tilde{\phi}_4&\hspace{-0.1cm}=&\hspace{-0.1cm}\frac{1}{4} \tanh ^{-1}\left[\frac{4 \sqrt{u}}{\sqrt{C_0^6 g^2 \kappa ^2 \left(C_6+2u\sigma g^3 \kappa ^3 \rho ^4\right)^2-16 C_5^2 \left(C_1^2+u\right)}}\right],\\
\phi_1&\hspace{-0.1cm}=&\hspace{-0.1cm}\frac{1}{4} \ln \left[C_1 \tanh{4\tilde{\phi}_3}+\sqrt{C_1^2\tanh ^2{4\tilde{\phi}_3}+u}\right]+\frac{1}{4} \ln \left[\frac{u\sigma g\kappa\rho}{4 \sqrt{2}}\right],\\
\phi_2&\hspace{-0.1cm}=&\hspace{-0.1cm}\frac{1}{4} \ln \left[C_1 \tanh{4\tilde{\phi}_3}+\sqrt{C_1^2\tanh ^2{4\tilde{\phi}_3}+u}\right]-\frac{1}{4} \ln \left[\frac{u\sigma g\kappa\rho}{4 \sqrt{2}}\right],\\
\varphi&\hspace{-0.1cm}=&\hspace{-0.1cm}\frac{3}{4}\ln\left[\sqrt{2} C_0 u\sigma g \kappa  \rho\right]+\frac{1}{4}\ln{\sinh{4\tilde{\phi}_4}},\\
A&\hspace{-0.1cm}=&\hspace{-0.1cm}\frac{1}{3}\varphi-\frac{1}{3}\ln{\sinh{4\tilde{\phi}_4}}
\end{eqnarray}
in which we have chosen the integration constants for $\phi_2$ and $A$ to be zero.
\\
\indent We end this section by pointing out that domain wall solutions obtained from gaugings in $\mathbf{56}^{+1}$ and $\mathbf{8}^{-3}$ representations can also be found by a similar analysis. The resulting solutions take the same form as the solutions given in this section with a sign change in some of the scalar fields. 
\section{Conclusions and discussions}\label{Discuss}
We have constructed the embedding tensors of six-dimensional maximal $N=(2,2)$ gauged supergravity for various gauge groups arising from the decomposition of the embedding tensor under $\mathbb{R}^+\times SO(4,4)\subset SO(5,5)$ symmetry. Under this decomposition, viable gauge groups can be determined from the embedding tensor in $\mathbf{8}^{\pm 1}$, $\mathbf{8}^{\pm 3}$, and $\mathbf{56}^{\pm 1}$ representations. We have pointed out that gaugings in $\mathbf{8}^{\pm 1}$ representation without $\mathbf{56}^{\pm 1}$ is not consistent due to the linear constraint, and gaugings in $\mathbf{8}^{\pm 3}$ representation only lead to a translational gauge group $\mathbb{R}^8$. On the other hand, gaugings in $\mathbf{56}^{\pm 1}$ representation give rise to $CSO(4-p,p,1)\sim SO(4-p,p)\ltimes \mathbb{R}^4$ gauge groups with $p=0,1,2$. Including $\mathbf{8}^{\pm 3}$ representation to $\mathbf{56}^{\mp 1}$ can enlarge the gauge groups to $SO(4-p,p+1)$ or $SO(5-p,p)$. We have also found a number of half-supersymmetric domain wall solutions from gaugings in $\mathbf{56}^{-1}$ and $\mathbf{8}^{+3}$ representations with various residual symmetries. The corresponding solutions for gaugings solely from $\mathbf{56}^{-1}$ representation can be straightforwardly obtained from these results by turning off the $\mathbf{8}^{+3}$ part. 
\\
\indent As pointed out in \cite{6D_Max_Gauging}, some of the gaugings under $SO(4,4)$ decomposition could be truncated to gaugings of half-maximal $N=(1,1)$ supergravity coupled to four vector multiplets in which supersymmetric AdS$_6$ vacua are known to exist \cite{AdS6_1,AdS6_2,AdS6_3}. It would be interesting to explicitly truncate the results given here to half-maximal gauged supergravity and obtain new gaugings as well as new supersymmetric $AdS_6$ vacua. Along this direction, a classification of gauge groups with known eleven-dimensional origins, arising from truncating the maximal theory to half-maximal one, has been given in \cite{6D_11}. It could also be interesting to extend this analysis to many new gaugings identified in this paper. 
\\
\indent On the other hand, uplifting the six-dimensional gauged supergravity and the corresponding domain wall solutions in this work and in \cite{our6DDW1,our6DDW2} to higher dimensions could also be worth considering. This could be done by constructing truncation ansatze of string/M-theory to six dimensions using $SO(5,5)$ exceptional field theory given in \cite{SO5_5_EFT1,SO5_5_EFT} and would lead to interesting holographic descriptions of maximal super Yang-Mills theory in five dimensions. Finally, finding a holographic interpretation of the domain wall solutions given in this paper could also be of particular interest. This could be done along the line of \cite{Spherical_Branes1} and \cite{Spherical_Branes2} in which a holographic description from a simple domain wall found in \cite{6D_Max_Cowdall} with $SO(5)$ symmetry has been studied.  
\vspace{0.5cm}\\
{\large{\textbf{Acknowledgement}}} \\
This work is supported by the National Research Council of Thailand (NRCT), Ramkhamhaeng University, and Chulalongkorn University under grants N42A660945 and N42A650263.
\appendix
\section{$SO(4,4)$ branching rules}\label{AppA}
In this appendix, we collect the decompositions for various representations of $SO(5,5)$ under $\mathbb{R}^+\times SO(4,4)$ in our convention. We note that these branching rules have also been given in \cite{6D_11}.

\subsection{Vector}\label{VecBranc}
Under $\mathbb{R}^+\times SO(4,4)\subset SO(5,5)$, the $SO(5,5)$ vector index is decomposed as $M=(-,I,+)$ where $I=1,...,8$ is an $SO(4,4)$ vector index. Accordingly, an $SO(5,5)$ vector $V_M$ can be written as
\begin{equation}
V_M=(V_-,V_I,V_+)\label{SO55_V_split}
\end{equation}
where $V_I$ is an $SO(4,4)$ vector. With the decomposition of the $SO(5,5)$ vector index $M=(-,I,+)$, the $SO(5,5)$ invariant metric together with its inverse are given by
\begin{equation}\label{off-diag-eta}
\eta_{MN}\ =\ \left(\begin{array}{c|c|c} 	 &  & 1 \\\hline
								 &\ \eta_{IJ\phantom{K}}  &  \\\hline
							1 &  &     \end{array}\right)\qquad\text{ and }\qquad\eta^{MN}\ =\ \left(\begin{array}{c|c|c} 	 &  & 1 \\\hline
								 & \ \eta^{IJ\phantom{\hat{K}}} &  \\\hline
							1 &  &     \end{array}\right).
\end{equation}
We have also introduced the $SO(4,4)$ invariant metric
\begin{equation}\label{SO(4,4)off-diag-eta}
\eta_{IJ}\ =\ \eta^{IJ}\ =\ \left(\begin{array}{c|c}  	 & \mathds{1}_4 \\\hline
							\mathds{1}_4 &     \end{array}\right)
\end{equation}
where $\mathds{1}_n$ denotes an $n\times n$ identity matrix. The $SO(4,4)$ vector index can be raised and lowered as $V^I=\eta^{IJ}V_J$ and $V_I=\eta_{IJ}V^J$.
\\
\indent The $SO(5,5)$ generators are defined in vector representation as
\begin{equation}
{(\boldsymbol{t}_{MN})_P}^Q\ =\ 4\eta_{P[M}\delta^Q_{N]}
\end{equation}
where $\delta^M_N=\mathds{1}_{10}$. These generators satisfy $\boldsymbol{t}_{MN}=\boldsymbol{t}_{[MN]}$ as well as the $SO(5,5)$ algebra 
\begin{equation}\label{SO(5,5)algebra}
\left[\boldsymbol{t}_{MN},\boldsymbol{t}_{PQ}\right]\ =\ 4(\eta_{M[P}\boldsymbol{t}_{Q]N}-\eta_{N[P}\boldsymbol{t}_{Q]M}).
\end{equation}
Under the $SO(4,4)$ decomposition, we define a generator corresponding to $\mathbb{R}^+\sim SO(1,1)\subset \mathbb{R}^+\times SO(4,4)$ as $\boldsymbol{d}\ =\ {\boldsymbol{t}}_{-+}$ whose explicit form in vector representation is
\begin{equation}
{\left(\boldsymbol{d}\right)_M}^N\ =\ \left(\begin{array}{c|c|c} -2	 &  &  \\\hline
								 & \phantom{-2} &  \\\hline
							 &  & 2    \end{array}\right).
\end{equation}
With $\boldsymbol{d}$ of this form, we can assign the $\mathbb{R}^+$ weights $\pm1$ to the two singlets of the $SO(4,4)$ in \eqref{SO55_V_split}. Therefore, the branching rule for a vector representation reads
\begin{equation}\label{VecDec}
\underbrace{\mathbf{10}}_{V_M}\ \rightarrow\ \underbrace{\mathbf{1}^{+1}}_{V_+}\,\oplus\,\underbrace{\mathbf{1}^{-1}}_{V_-}\,\oplus\,\underbrace{\mathbf{8}^{\,0}}_{V_I}\, .
\end{equation}
\subsection{Adjoint}\label{AdjBranc}
The decomposition of adjoint representation follows from the branching rule of vector representations. Using $M=(-,I,+)$, we can decompose the $SO(5,5)$ generators as 
\begin{equation}
\boldsymbol{t}_{MN}\ =\ \left(\begin{array}{c|c|c} 	 & \boldsymbol{k}_{J} & \boldsymbol{d} \\\hline
								-\boldsymbol{k}_{I} & \boldsymbol{\tau}_{IJ} & -\boldsymbol{p}_{I} \\\hline
							-\boldsymbol{d} & \boldsymbol{p}_{J} &     \end{array}\right)\label{SO5_5_gen_decom}
\end{equation}
where we have identified $\boldsymbol{t}_{+I}=\boldsymbol{p}_{I}$ and $\boldsymbol{t}_{-I}=\boldsymbol{k}_{I}$.
From the $SO(5,5)$ algebra \eqref{SO(5,5)algebra}, we can derive the following commutation relations 
\begin{eqnarray}
\left[\boldsymbol{d},\boldsymbol{d}\right]&=&0,\qquad\qquad\quad \left[\boldsymbol{d},\boldsymbol{\tau}_{IJ}\right]\ =\ 0,\label{R+ComRel1}\\
\left[\boldsymbol{d},\boldsymbol{p}_{I}\right]&=&-2\boldsymbol{p}_{I},\quad\quad\quad\ \,\, \left[\boldsymbol{d},\boldsymbol{k}_{I}\right]\ =\ +2\boldsymbol{k}_{I},\label{R+ComRel2}\\
\left[\boldsymbol{p}_{I},\boldsymbol{p}_{J}\right]&=&0,\qquad\quad\qquad \left[\boldsymbol{k}_{I},\boldsymbol{k}_{J}\right]\ =\ 0,\\
\left[\boldsymbol{p}_{I},\boldsymbol{\tau}_{JK}\right]&=&4\eta_{I[J}\boldsymbol{p}_{K]},\quad\ \left[\boldsymbol{k}_{I},\boldsymbol{\tau}_{JK}\right]\ =\ 4\eta_{I[J}\boldsymbol{k}_{K]},\\
\left[\boldsymbol{p}_{I},\boldsymbol{k}_{J}\right]&=&2(\boldsymbol{d}\eta_{IJ}-\boldsymbol{\tau}_{IJ}),\\
\left[\boldsymbol{\tau}_{IJ},\boldsymbol{\tau}_{KL}\right]&=&4(\eta_{I[K}\boldsymbol{\tau}_{L]J}-\eta_{J[K}\boldsymbol{\tau}_{L]I}).\label{R+ComRelLast}
\end{eqnarray}
The last commutation relation is the $SO(4,4)$ algebra implying that $\boldsymbol{\tau}_{IJ}$ are $SO(4,4)$ generators. It follows that the $SO(4,4)$ branching rule for the adjoint representation of $SO(5,5)$ is given by
\begin{equation}\label{SO55GenDec}
\underbrace{\mathbf{45}}_{\boldsymbol{t}_{MN}}\ \rightarrow\ \underbrace{\mathbf{1}^{0}}_{\boldsymbol{d}}\,\oplus\,\underbrace{\mathbf{28}^{0}}_{\boldsymbol{\tau}_{IJ}}\,\oplus\,\underbrace{\mathbf{8}^{+2}}_{\boldsymbol{k}_{I}}\,\oplus\,\underbrace{\mathbf{8}^{-2}}_{\boldsymbol{p}_I}
\end{equation}
where the $\mathbb{R}^+$ weights are determined from the commutation relations given in \eqref{R+ComRel1} and \eqref{R+ComRel2}.

\subsection{Spinor}\label{SpinorBranc}
In order to find the $SO(4,4)$ branching rule for spinor representation, we start from the 32-dimensional $SO(5,5)$ gamma matrices given in \cite{6D_11}
\begin{eqnarray}
\tilde{\mathbf{\Gamma}}_{\underline{1}}&=&\sigma_1\otimes\sigma_3\otimes\mathds{1}_2\otimes\mathds{1}_2\otimes\mathds{1}_2,\qquad\tilde{\mathbf{\Gamma}}_{\underline{2}}\ =\ \sigma_1\otimes\sigma_1\otimes\sigma_2\otimes\sigma_1\otimes\sigma_2,\nonumber\\
\tilde{\mathbf{\Gamma}}_{\underline{3}}&=&\sigma_1\otimes\sigma_1\otimes\sigma_1\otimes\sigma_1\otimes\mathds{1}_2,\qquad\tilde{\mathbf{\Gamma}}_{\underline{4}}\ =\ \sigma_1\otimes\sigma_1\otimes\sigma_1\otimes\sigma_3\otimes\sigma_1,\nonumber\\\tilde{\mathbf{\Gamma}}_{\underline{5}}&=&\sigma_1\otimes\sigma_1\otimes\sigma_1\otimes\sigma_3\otimes\sigma_3,\qquad\tilde{\mathbf{\Gamma}}_{\underline{6}}\ =\ i \sigma_1\otimes\sigma_1\otimes\sigma_3\otimes\sigma_1\otimes\sigma_2,\nonumber\\
\tilde{\mathbf{\Gamma}}_{\underline{7}}&=&i\sigma_1\otimes\sigma_1\otimes\mathds{1}_2\otimes\sigma_2\otimes\mathds{1}_2,\quad\ \ \tilde{\mathbf{\Gamma}}_{\underline{8}}\ =\ i\sigma_1\otimes\sigma_1\otimes\mathds{1}_2\otimes\sigma_3\otimes\sigma_2,\nonumber\\
\tilde{\mathbf{\Gamma}}_{\underline{9}}&=&i\sigma_1\otimes\sigma_2\otimes\mathds{1}_2\otimes\mathds{1}_2\otimes\mathds{1}_2,\quad\ \ \tilde{\mathbf{\Gamma}}_{\underline{10}}=\ i\sigma_2\otimes\mathds{1}_2\otimes\mathds{1}_2\otimes\mathds{1}_2\otimes\mathds{1}_2\quad\label{ExplicitDiagGamma}
\end{eqnarray}
where $\{\sigma_1,\sigma_2,\sigma_3\}$ are the usual Pauli matrices
\begin{equation}
\sigma_1\ = \ \begin{pmatrix} 0 & 1\\ 1 & 0 \end{pmatrix},\qquad \sigma_2\ = \ \begin{pmatrix} 0 & -i\\ i & 0 \end{pmatrix},\qquad\sigma_3\ = \ \begin{pmatrix} 1 & 0\\ 0 & -1 \end{pmatrix}.
\end{equation}
The $SO(5,5)$ gamma matrices $\tilde{\mathbf{\Gamma}}_{\underline{A}}$ with $\underline{A}=\underline{1},...,\underline{10}$ satisfy the Clifford algebra
\begin{equation}
\left\{\tilde{\mathbf{\Gamma}}_{\underline{A}},\tilde{\mathbf{\Gamma}}_{\underline{B}}\right\}\ =\ 2 \eta_{\underline{A}\underline{B}}\mathds{1}_{32}
\end{equation}
in which $\eta_{\underline{A}\underline{B}} = \textrm{diag}(\mathds{1}_5,-\mathds{1}_5)$ is the $SO(5,5)$ invariant metric in diagonal basis. In this explicit representation, the $SO(5,5)$ chirality matrix is given by
\begin{equation}
\tilde{\mathbf{\Gamma}}_\ast=\tilde{\mathbf{\Gamma}}_{\underline{1}...\underline{10}}=\textrm{diag}(-\mathds{1}_{16},\mathds{1}_{16})
\end{equation}
and the $SO(5,5)$ charge conjugation matrix is
\begin{equation}\label{CCDef}
\tilde{\mathds{C}}\ =\ i \sigma_2\otimes\sigma_3\otimes\sigma_3\otimes\mathds{1}_2\otimes\mathds{1}_2
\end{equation}
satisfying
\begin{equation}
\tilde{\mathbf{\Gamma}}_{\underline{A}}\tilde{\mathds{C}}=(\tilde{\mathbf{\Gamma}}_{\underline{A}}\tilde{\mathds{C}})^T\qquad\text{ and }\qquad \tilde{\mathds{C}}^T=-\tilde{\mathds{C}}.
\end{equation}
\indent We now transform all these results to the basis with off-diagonal metric $\eta_{MN}$ given in \eqref{off-diag-eta}. In this basis, the Clifford algebra reads
\begin{equation}
\left\{\tilde{\mathbf{\Gamma}}_M,\tilde{\mathbf{\Gamma}}_N\right\}\ =\ 2 \eta_{MN}\mathds{1}_{32}\, .
\end{equation}
The 32-dimensional $SO(5,5)$ gamma matrices in the off-diagonal basis can be obtained from 
\begin{equation}
\tilde{\mathbf{\Gamma}}_M\ =\ {\mathbb{M}_M}^{\underline{A}}\tilde{\mathbf{\Gamma}}_{\underline{A}}
\end{equation}
with the transformation matrix
\begin{equation}\label{offDiagTrans}
\mathbb{M}\ = \ \frac{1}{\sqrt{2}}(\eta_{\text{diag}}+\eta_{\text{off-diag}})\ =\ \frac{1}{\sqrt{2}}\left(\begin{array}{c|c|c|c} 1 &  &  & 1\\\hline  & \mathds{1}_4 & \mathds{1}_4 &  \\\hline  & \mathds{1}_4 & -\mathds{1}_4 &  \\\hline 1 &  &  & -1\end{array}\right).
\end{equation}
Using this transformation matrix, the relation between diagonal and off-diagonal $\eta$ is found to be 
\begin{equation}
\eta_{MN}\ =\ {\mathbb{M}_M}^{\underline{A}}\,{\mathbb{M}_N}^{\underline{B}}\,\eta_{\underline{A}\underline{B}}.
\end{equation}
\indent The $SO(5,5)$ gamma matrices are chirally decomposed as
\begin{equation}\label{SO(5,5)Dec}
{(\tilde{\mathbf{\Gamma}}_M)_{\mathcal{A}}}^{\mathcal{B}}\ =\ \left(\begin{array}{c|c}  	 & {(\Gamma_M)_{A}}^{B'} \\\hline
							{(\Gamma_M)_{A'}}^{B\phantom{\hat{i}}} &     \end{array}\right).
\end{equation}
with spinor index $\mathcal{A}=1,2,...,32$ split according to $\mathcal{A}=(A,A')$  where $A=1,...,16$ and $A'=17,...,32$. We can raise and lower the chirally decomposed spinor indices with the charge conjugation matrix given in \eqref{CCDef} such that
\begin{equation}
(\tilde{\mathbf{\Gamma}}_M)_{\mathcal{A}\mathcal{B}}\ =\ {(\tilde{\mathbf{\Gamma}}_M)_{\mathcal{A}}}^{\mathcal{C}}\tilde{\mathds{C}}_{\mathcal{C}\mathcal{B}}\ =\ \left(\begin{array}{c|c}   (\Gamma_M)_{AB} &  \\\hline
							 &  (\Gamma_M)_{A'B'}   \end{array}\right).
\end{equation}
The chirally decomposed $16\times16$ gamma matrices $(\Gamma_M)_{AB}$ play an important role in determining explicit forms of the embedding tensor. 
\\
\indent Following \cite{6D_11}, $SO(4,4)$ gamma matrices can be extracted from the $SO(5,5)$ ones using the decomposition $M=(-,I,+)$. The ten $32$-dimensional $SO(5,5)$ gamma matrices are also decomposed as $\tilde{\mathbf{\Gamma}}_M=(\tilde{\mathbf{\Gamma}}_-,\tilde{\mathbf{\Gamma}}_I,\tilde{\mathbf{\Gamma}}_+)$. In the explicit representation \eqref{ExplicitDiagGamma}, only $\tilde{\mathbf{\Gamma}}_I$ can be written as
\begin{equation}
\tilde{\mathbf{\Gamma}}_I=\sigma_1\otimes\Gamma_I
\end{equation}
so that the chirally decomposed $16\times16$ gamma matrices on the upper-right block and the lower-left block in \eqref{SO(5,5)Dec} are the same, i.e., ${(\Gamma_I)_{A}}^{B'}={(\Gamma_I)_{A'}}^{B}$. We can redefine these $16\times16$ gamma matrices to be the $SO(4,4)$ gamma matrices ${(\Gamma_I)_{A}}^{B}$ satisfying the Clifford algebra
\begin{equation}
\{\Gamma_I,\Gamma_J\}=2\eta_{IJ}\mathds{1}_{16}.
\end{equation}
In this case, the $SO(4,4)$ chirality matrix is 
\begin{equation}
\Gamma_9=\text{diag}(\mathds{1}_8,-\mathds{1}_8),
\end{equation}
and the $SO(4,4)$ charge conjugation matrix, satisfying $\Gamma_I\mathds{C}=(\Gamma_I\mathds{C})^T$ and $\mathds{C}^T=\mathds{C}$, is given by
\begin{equation}
\mathds{C}=\sigma_3\otimes\sigma_3\otimes\mathds{1}_2\otimes\mathds{1}_2.
\end{equation}
\indent Decomposing the spinor index $A=(m,\dot{m})$ where $m=1,...,8$ and $\dot{m}=\dot{1},...,\dot{8}$, we can write this charge conjugation matrix in the form
\begin{equation}
\mathds{C}_{AB}\ =\ \left(\begin{array}{c|c}   \mathbf{c}_{mn} &  \\\hline
							 &  \mathbf{c}_{\dot{m}\dot{n}}   \end{array}\right)\quad\text{ and }\quad\mathds{C}^{AB}\ =\ \left(\begin{array}{c|c}   \mathbf{c}^{mn} &  \\\hline
							 &  \mathbf{c}^{\dot{m}\dot{n}}   \end{array}\right)\label{SO4_4_CC}
\end{equation}
where $\mathbf{c}_{mn}=\mathbf{c}^{mn}=-\mathbf{c}_{\dot{m}\dot{n}}=-\mathbf{c}^{\dot{m}\dot{n}}=\text{diag}(\mathds{1}_4,-\mathds{1}_4)$. Besides, we find that the $SO(4,4)$ gamma matrices are also chirally decomposed as
\begin{equation}\label{gammaIdecom1}
{(\Gamma_I)_A}^B\ =\ \left(\begin{array}{c|c}  	 & {(\gamma_I)_{m}}^{\dot{n}} \\\hline
							{(\gamma_I)_{\dot{m}}}^{n\phantom{'}} &     \end{array}\right)
\end{equation}
in which ${(\gamma_I)_{m}}^{\dot{n}}$ and ${(\gamma_I)_{\dot{m}}}^{n}$ are in turn chirally decomposed $8\times8$ gamma matrices. The spinor index $B$ can be lowered using the $SO(4,4)$ charge conjugation matrix as
\begin{equation}\label{gammaIdecom2}
(\Gamma_I)_{AB}\ =\ {(\Gamma_I)_A}^{C}\mathds{C}_{CB}\ =\ \left(\begin{array}{c|c}    & (\gamma_I)_{m\dot{n}} \\\hline
							(\gamma_I)_{\dot{m}n} &     \end{array}\right)
\end{equation}
where $(\gamma_I)_{\dot{n}m}= (\gamma_I)_{m\dot{n}}$ due to $(\Gamma_M)_{BA}=(\Gamma_M)_{AB}$. 
\\
\indent Following the same decomposition of spinor indices, we find the remaining two $SO(5,5)$ gamma matrices of the forms
\begin{equation}
(\Gamma_-)_{AB}=-\sqrt{2}\left(\begin{array}{c|c}  \mathbf{c}_{mn} & \phantom{\mathds{1}_8}   \\\hline
							 &     \end{array}\right)\qquad\text{ and }\qquad
(\Gamma_+)_{AB}=\sqrt{2}\left(\begin{array}{c|c}  \phantom{\mathds{1}_8} &    \\\hline
							 &  \mathbf{c}_{\dot{m}\dot{n}}   \end{array}\right).
\end{equation}
The $SO(5,5)$ generators in spinor representation satisfying \eqref{SO(5,5)algebra} are given by
\begin{equation}\label{SO55GenSpin}
{(\boldsymbol{t}_{MN})_A}^B\ =\ {(\Gamma_{MN})_A}^B
\end{equation}
where
\begin{equation}
{(\Gamma_{MN})_A}^B=\frac{1}{2}\left[{(\Gamma_M)_A}^{C'}{(\Gamma_N)_{C'}}^B-{(\Gamma_N)_A}^{C'}{(\Gamma_M)_{C'}}^B\right].
\end{equation}
It should be noted that the $SO(5,5)$ generators in spinor representation given in \eqref{SO55GenSpin} also decompose according to \eqref{SO55GenDec} and satisfy the same commutation relations given in \eqref{R+ComRel1} to \eqref{R+ComRelLast}. In particular, with the explicit representation \eqref{ExplicitDiagGamma}, the $\mathbb{R}^+$ generator in spinor representation is 
\begin{equation}
{\left(\boldsymbol{d}\right)_A}^B\ =\ \left(\begin{array}{c|c}   \mathds{1}_8 &  \\\hline
							 &  -\mathds{1}_8  \end{array}\right).
\end{equation}
\indent The $SO(4,4)$ branching rule for spinor representation is now straightforward. For a given $SO(5,5)$ spinor in $\mathbf{16}_s$ representation $\Psi_A$, we have the decomposition
\begin{equation}
\Psi_A\ =\ \left(\begin{array}{c}   \Psi_m \\
					   \Psi_{\dot{m}}   \end{array}\right).
\end{equation}
By assigning the $\mathbb{R}^+$ weights $+1$ and $-1$ to $\Psi_m$ and $\Psi_{\dot{m}}$, we obtain the branching rule of the form
\begin{equation}
\underbrace{\mathbf{16}_s}_{\Psi_A}\ \rightarrow\ \underbrace{\mathbf{8}^{+1}}_{\Psi_m}\,\oplus\,\underbrace{\mathbf{8}^{-1}}_{\Psi_{\dot{m}}}\, .
\end{equation}

\subsection{Vector-spinor}\label{Apptheta}
The vector-spinor representation of $SO(5,5)$ we are interested in is given by a $16\times10$ matrix $\theta^{AM}\in\mathbf{144}_c$ subject to the linear constraint \eqref{MainLC}, $(\Gamma_M)_{AB}\,\theta^{BM}\ =\ 0\,$. This linear constraint reduces $160$ components of $\theta^{AM}$ to $144$ in $\mathbf{144}_c$ representation. 
\\
\indent With the decomposition of $SO(5,5)$ vector and spinor indices as $M=(-,I,+)$ and $A=(m,\dot{m})$, we decompose the $\theta^{AM}$ matrix as
\begin{equation}
\theta^{AM}\ = \ \left(\begin{array}{c|c|c} \theta_1^{m -}  & \theta_3^{m I} & \theta_5^{m +} \\\hline
								\theta_2^{\dot{m} -} & \theta_4^{\dot{m} I} & \theta_6^{\dot{m} +\phantom{'}} \end{array}\right).
\end{equation}
It is straightforward to determine that $\mathbb{R}^+$ weights of $(\theta_1^{m -},\theta_2^{\dot{m} -},\theta_3^{m I},\theta_4^{\dot{m} I},\theta_5^{m +},\theta_6^{\dot{m}})$ are given by $(+1,+3,-1,+1,-3,-1)$. Finally, as in \cite{6D_Max_Gauging}, the $SO(4,4)$ branching rule for $\mathbf{144}_c$ representation is given by 
\begin{equation}\label{thetaDec}
\underbrace{\mathbf{144}_c}_{\theta^{AM}}\ \rightarrow\ \underbrace{\mathbf{56}^{-1}}_{\vartheta_3^{m I}}\,\oplus\,\underbrace{\mathbf{56}^{+1}}_{\vartheta_4^{\dot{m} I}}\,\oplus\,\underbrace{\mathbf{8}^{-1}}_{\theta_6^{\dot{m} +}}\,\oplus\,\underbrace{\mathbf{8}^{+1}}_{\theta_1^{m -}}\,\oplus\,\underbrace{\mathbf{8}^{+3}}_{\theta_2^{\dot{m} -}}\,\oplus\,\underbrace{\mathbf{8}^{-3}}_{\theta_5^{m +}}
\end{equation}
with
\begin{equation}
\vartheta_3^{m I}=\theta_3^{m I}+\frac{\sqrt{2}}{8}{(\gamma^I)^{m}}_{\dot{n}}\theta_6^{\dot{n} +}\qquad\text{and }\qquad\vartheta_4^{\dot{m} I}=\theta_4^{\dot{m} I}-\frac{\sqrt{2}}{8}{(\gamma^I)^{\dot{m}}}_n\theta_1^{n -}\, .\label{vartheta_def}
\end{equation}
These two definitions follow from the linear constraint decomposed under $SO(4,4)$ as
\begin{equation}\label{red_LC}
0=\sqrt{2}\mathbf{c}_{mn}\theta_1^{n -}-(\gamma_I)_{m\dot{n}}\theta_4^{\dot{n} I}\qquad\text{ and }\qquad 0=\sqrt{2}\mathbf{c}_{\dot{m}\dot{n}}\theta_6^{\dot{n} +}+(\gamma_I)_{\dot{m}n}\theta_3^{n I}\, .
\end{equation}
It is also useful to note that $\vartheta_3^{m I}$ and $\vartheta_4^{\dot{m} I}$ satisfy the following conditions 
\begin{equation}\label{var_red_LC}
(\gamma_I)_{\dot{m}n}\vartheta_3^{n I}=0\qquad\text{ and }\qquad(\gamma_I)_{m\dot{n}}\vartheta_4^{\dot{n} I}=0\, .
\end{equation}

\end{document}